\documentclass{jfm}
\usepackage{graphicx}
\usepackage{epstopdf, epsfig}
\usepackage{color}
\usepackage{tasks}
\usepackage{array}
\usepackage{tabularx}
\usepackage{amssymb}

\newcommand{\be}{\begin{eqnarray}}
\newcommand{\ee}{\end{eqnarray}}
\newcommand{\bse}{\begin{subequations}}
\newcommand{\ese}{\end{subequations}}

\newcommand{\bnum}{\begin{enumerate}}
\newcommand{\enum}{\end{enumerate}}

\newcommand{\bit}{\begin{itemize}}
\newcommand{\eit}{\end{itemize}}

\newcommand{\bc}{\begin{cases}}
\newcommand{\ec}{\end{cases}}

\newcommand{\bpm}{\begin{pmatrix}}
\newcommand{\epm}{\end{pmatrix}}

\newcommand{\bvm}{\begin{vmatrix}}
\newcommand{\evm}{\end{vmatrix}}

\newcommand{\bs}{\boldsymbol}

\newcommand{\mcal}{\mathcal}

\newcommand{\vol}{\mrm{vol}}

\newcommand{\eps}{\epsilon}

\newcommand{\gk}{\kappa}

\newcommand{\gs}{\sigma}

\newcommand{\Gl}{\Lambda}

\newcommand{\f}{\frac}

\newcommand{\tn}{\textnormal}

\shorttitle{Triad interaction in active turbulence}
\shortauthor{J. S\l{}omka, P. Suwara and J. Dunkel}

\title{The nature of triad interactions in active turbulence}

\author{Jonasz S\l{}omka\aff{1},
 Piotr Suwara\aff{1}, 
 \and
  J{\"o}rn Dunkel\aff{1}\corresp{\email{dunkel@mit.edu}}
}

\affiliation{\aff{1}Department of Mathematics, Massachusetts Institute of Technology, 77 Massachusetts Avenue, Cambridge, MA 02139-4307, USA}

\begin{document}

\maketitle

\begin{abstract}
Generalized Navier-Stokes (GNS) equations describing three-dimensional (3D) active fluids with flow-dependent spectral forcing have been shown to possess   numerical solutions that can sustain significant  energy transfer to larger scales by realising chiral Beltrami-type chaotic flows. To rationalise these findings, we study here the triad truncations of polynomial and Gaussian GNS models focusing on modes lying in the energy injection range. Identifying a previously unknown cubic invariant, we show that the asymptotic triad dynamics reduces to  that of a forced rigid body coupled to a particle moving in a magnetic field. This analogy allows us to classify triadic interactions by their asymptotic stability: unstable triads correspond to rigid-body forcing along the largest and smallest principal axes, whereas stable triads arise from forcing along the middle axis. Analysis of the polynomial GNS model reveals that unstable triads induce exponential growth of energy and helicity, whereas stable triads develop a limit cycle of bounded energy and helicity. This suggests that the unstable triads dominate the initial relaxation stage of the full hydrodynamic equations,  whereas the stable triads determine the statistically stationary state. To test this hypothesis, we introduce and investigate the Gaussian active turbulence model, which develops a Kolmogorov-type $-5/3$ energy spectrum at large wavelengths. Similar to the polynomial case, the steady-state chaotic flows spontaneously accumulate non-zero mean helicity while exhibiting Beltrami statistics and upward energy transport. Our results suggest that  self-sustained Beltrami-type flows and an inverse energy cascade may be generic features of 3D active turbulence models with flow-dependent  spectral forcing.
\end{abstract}


\section{Introduction}

Originally introduced by~\citet{kraichnan1973helical} to study energy transfer in inertial turbulence, the triad truncation projects the fluid dynamics onto three Fourier modes with wavevectors $\{\bs k,\bs p,\bs q\}$ such that $\bs k+\bs p+\bs q=\bs 0$. The truncated dynamics of isolated triads differs from the exact fluid flow, failing for example to conserve the topology of the vorticity field~\citep{moffatt2014note}. Notwithstanding, the analysis of triadic interactions has yielded important qualitative insights about the direction of energy transfer in externally forced~\citep{waleffe1992,waleffe1993inertial} and magnetohydrodynamic~\citep{lessinnes2009helical,linkmann2016helical,linkmann2017triad} turbulence.~\citet{kraichnan1973helical} combined the triad truncation with absolute equilibrium considerations to argue against the possibility of an inverse inertial energy cascade in  three-dimensional (3D) helical turbulence~\citep{brissaud1973helicity}. Direct numerical simulations of the Navier--Stokes equations (NS) verified later that such turbulence indeed produces only direct energy and helicity cascades~\citep{borue1997spectra}. In the meantime,~\citet{waleffe1992,waleffe1993inertial} formulated his instability assumption, suggesting that there exists  a subclass of triads capable of transferring energy to larger scales, but that this subclass is not dominant in isotropic and reflection-invariant turbulence. To amplify the impact of such upward-cascading triads, \citet{biferale2012inverse,biferale2013split} studied a projection of  the NS equations onto positive helicity states, which breaks reflection-invariance and eliminates triads promoting forward energy transfer, and found that inverse energy transfer can develop in such a reduced system. Similar conclusions apply to NS-like equations where the nonlinear term is modified to weight various types of triadic interactions differently~\citep{sahoo2017discontinuous}. New analytical properties of the triadic system continue to be discovered, including pseudo-invariants for a subclass of the interactions~\citep{RathmannDitlevsen2017}, with direct implications for externally driven turbulence in passive fluids. 
\par
Building on work by~\citet{moffatt2014note}, we will extend here the analysis of triad truncations to a class of generalized Navier-Stokes (GNS) equations that constitute effective phenomenological models~\citep{2017SlomkaDunkel_PRF,2017SlomkaDunkel} for intrinsically driven chaotic flows in active fluids~\citep{2008SaintillanShelley,2013Marchetti_Review,2013Lauga,2015Giomi}, arising from the non-equilibrium stresses exerted by biological or engineered active components~\citep{mendelson1999organized,dombrowski2004self,howse2007self,2008Walther_SM}. The recent numerical investigation of a polynomial GNS model~\citep{2017SlomkaDunkel} suggested that active suspensions, such as water-based solutions driven by swimming bacteria~\citep{2007SoEtAl,Dunkel2013_PRL} or micro-tubule networks~\citep{2012Sanchez_Nature}, can spontaneously break mirror-symmetry and develop upward energy transfer even in 3D. The analysis below rationalizes these findings by identifying a previously unknown cubic invariant, which allows us to classify and contrast the triad dynamics for the classical Euler and the GNS equations.  For the GNS case, we show that the asymptotic dynamics reduces to  that of a forced rigid body coupled to a particle moving in a magnetic field. For the classical Euler triads, we combine the cubic invariant with the conservation of in-plane energy and enstrophy~\citep{moffatt2014note} to characterise in detail the geometry of the solution space.

\subsection{Generalized Navier-Stokes equations for active turbulence}
Classical turbulence concerns externally driven flows at high Reynolds number~\citep{2004Frisch}. By contrast, energy injection in suspensions of self-motile structures~\citep{NeedlemanDogic2017} is delocalised and inherently coupled to the flow field. For example, swimming microorganisms~\citep{mendelson1999organized,dombrowski2004self,2010Pedley,ishikawa2011energy,Dunkel2013_PRL} stir the surrounding fluid, but also respond to the flow field and interact through the fluid. Similar flow-dependent forcing mechanisms are present in suspensions of artificial micro-swimmers~\citep{howse2007self,2008Walther_SM,Bricard:2013aa} or ATP-driven microtubule networks~\citep{2012Sanchez_Nature}. When the concentration of such active objects is sufficiently high, self-sustained chaotic flow patterns emerge; this phenomenon is commonly referred to as active turbulence nowadays~\citep{wolgemuth2008collective,2012Wensink,2015Giomi,Bratanov08122015,urzay2017multi}. A striking difference between classical and active turbulence is that the latter often exhibits characteristic scales, leading to a preferred eddie size~\citep{2012Sokolov,2007SoEtAl,2012Wensink,Dunkel2013_PRL,2012Sanchez_Nature}. A minimal phenomenological model combing scale selection with flow-dependent driving is given by the higher-order GNS equations~\citep{2017SlomkaDunkel_PRF,2017SlomkaDunkel}
\bse
\label{e:eom}
\be
\label{e:eoma}
\nabla\cdot\bs v&=&0, \\
\label{e:eomb}
\p_t \bs v+\bs v \cdot \nabla \bs v&=&-\nabla p+\nabla \cdot \mathsfbi{\boldsymbol\gs},
\ee
\ese
where the higher-order stress tensor
\be
\label{e:poly_stress}
\mathsfbi{\boldsymbol\gs}=(\Gamma_0 -\Gamma_2 \nabla^2+\Gamma_4 \nabla^4)[\nabla \bs v+ (\nabla \bs v)^\top],
\ee
with $\nabla^{2n}\equiv (\nabla^2)^n$, \mbox{$n\ge2$}, accounts effectively for both passive contributions from the intrinsic solvent fluid viscosity and active contributions representing the stresses exerted by the microswimmers on the fluid.  Related higher-order Navier--Stokes models have been studied previously in the context of soft-mode turbulence and seismic waves~\citep{1993BeNi_PhysD,1996Tribelsky_PRL,PhysRevE.77.035202} so that the considerations below may extend to these systems as well. On a periodic cubic domain, the Fourier representation of~(\ref{e:eom}) and (\ref{e:poly_stress}) reads
\begin{figure}
  \centerline{\includegraphics[width=0.9\textwidth]{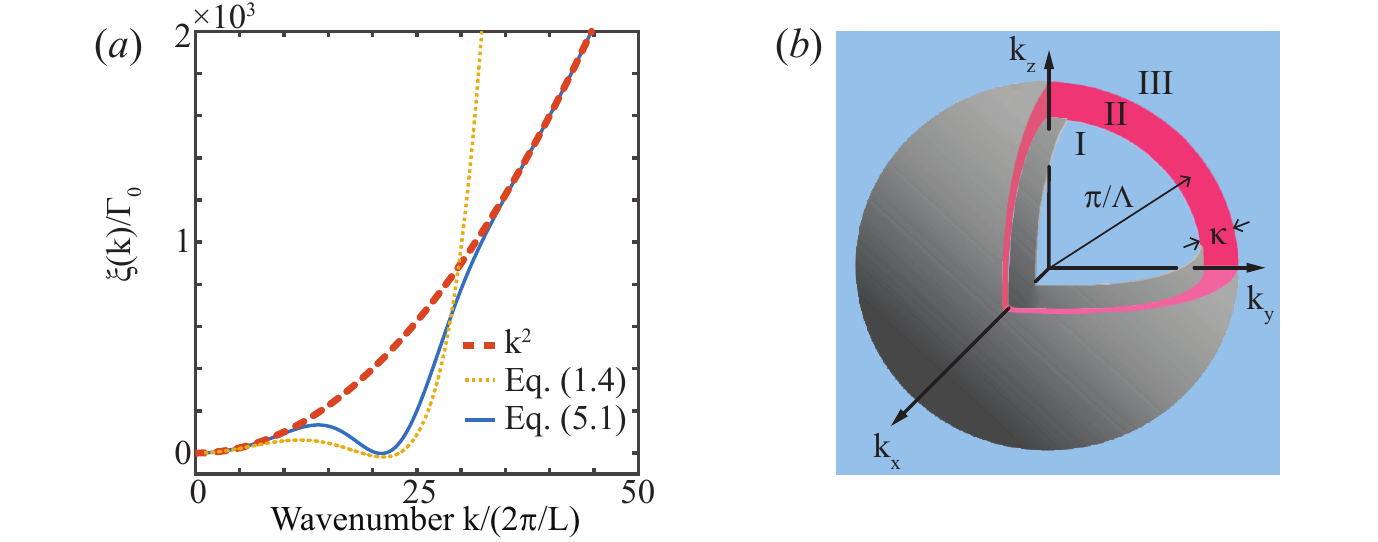}}
  \caption{
(\textit{a})~Dispersion relations $\xi(k)$ for the polynomial GNS model~(\ref{e:dispersion_relation_higher_order}) and the Gaussian GNS model~(\ref{e:xi_newmodel}). Modes with $\xi(k)<0$ define the energy injection and scale selection domain typical of active turbulence.
(\textit{b})~Ilustration of the key model parameters in 3D Fourier space. 
The spectral bandwidth $\kappa$ defines the width of the unstable domain II (red), which is localized around the characterized vortex scale $\Lambda$ and separates  dissipative Fourier modes at large (region I) and smale scales (region III);  reproduced with permission from~\citep{2017SlomkaDunkel}.
  }
\label{fig:dispersion_relations}
\end{figure}
\be
\label{e:eom_fourier}
\Big[\f{\p}{\p t}+\xi(k)\Big]\hat{v}_i(\bs k,t)=-i\sum_{\bs k+\bs p+\bs q=\bs 0}P_{ij}(\bs k)
q_k \hat{v}_k^*(\bs p,t)\hat{v}_j^*(\bs q,t),
\ee
where $k=|\bs k|$, the projector $P_{ij}=\delta_{ij}-k_ik_j/k^2$ enforces incompressibility, and the dispersion relation is given by the polynomial
\be
\label{e:dispersion_relation_higher_order}
\xi(k)=\Gamma_0 k^2+\Gamma_2 k^4+\Gamma_4 k^6,
\ee
see figure~\ref{fig:dispersion_relations}(\textit{a}). Microswimmer activity is modelled by letting $\Gamma_2<0$, which introduces a band of linearly unstable modes with $\xi(k)<0$, while $\Gamma_0>0$ and $\Gamma_4>0$ represent damping at large and small scales with $\xi(k)>0$. The most unstable wavenumber $k_\Lambda$ determines the typical eddy size $\Lambda=\pi/k_\Lambda$, the corresponding growth rate sets the timescale $\tau=-\xi(k_\Lambda)^{-1}$, and we denote by $\gk$ the bandwidth of the unstable modes, see figure~\ref{fig:dispersion_relations}(\textit{b}). The parameters $(\Lambda,\tau,\kappa)$, uniquely determined by $(\Gamma_0,\Gamma_2,\Gamma_4)$, characterise the resulting flow structures and can be inferred from experimental data~\citep{2017SlomkaDunkel}. Numerical simulations showed that the polynomial GNS model defined by~(\ref{e:eom_fourier}) and (\ref{e:dispersion_relation_higher_order}) exhibits  spontaneous mirror symmetry breaking by developing helical flow structures that are statistically close to Beltrami fields. The strength of the symmetry breaking is controlled by the active bandwidth $\kappa$, corresponding to the red domain in figure~\ref{fig:dispersion_relations}(\textit{b}). For sufficiently small $\kappa \ll \Gl^{-1}$, an upward energy transfer develops in an otherwise homogeneous and isotropic active turbulence~\citep{2017SlomkaDunkel}.

\subsection{Triad interactions in active turbulence}
 
 In this work, we investigate analytically and numerically the dynamical system arising from the triad truncation of~(\ref{e:eom_fourier}). In contrast to the approach typically adopted when studying the inertial energy transfer in classical turbulence, our analysis does \emph{not} neglect the linear term $\xi(k)$, although we will later discuss the implications for the classical case $\xi(k)\equiv 0$ as well. Specifically, we focus on the subclass of all possible triad interactions in which one or two \lq legs\rq~lie in the energy injection range, while the remaining legs are dissipative. We refer to such triads as \lq active triads\rq, to distinguish them from the \lq classical triads\rq~for which $\xi(k)\equiv 0$. Utilizing a previously unrecognized cubic invariant, we show that the resulting triad dynamics is asymptotically equivalent to a coupled system of a rigid body and a particle moving in a magnetic field. This analogy allows us to classify the active triads by their asymptotic stability: Triads forced at the small or large scale are unstable and increase energy and helicity exponentially, whereas triads forced at the intermediate scale are stable and develop a limit cycle. This asymptotic behaviour of the active triads is in stark contrast to the classical triadic dynamics, for which the rigid body analogy does \emph{not} hold in general but whose solutions one can classify using the cubic invariant. For the untruncated system~(\ref{e:eom_fourier}), it is plausible that unstable active triads dominate the initial relaxation characterised by helicity growth, whereas stable active triads determine the subsequent statistically stationary stage. To support this hypothesis, we will also consider a non-polynomial active turbulence model~(\ref{e:xi_newmodel}) which combines the usual viscous dissipation $\sim \Gamma_0 k^2$ with a  Gaussian forcing term, see blue solid curve in figure~\ref{fig:dispersion_relations}(\textit{a}). We will use direct numerical simulations to show that the Gaussian activity model develops steady-state energy spectra that approximately follow the Kolmogorov $-5/3$ scaling~\citep{kolmogorov1941local} at large wavelengths. The steady-state velocity and vorticity fields become strongly aligned and the upward energy transfer is balanced by viscous dissipation. These results suggest that Beltrami-type flows and an inverse energy cascade are generic features of 3D active turbulence models with flow-dependent spectral forcing.
\vspace{0.2cm}

\section{Triad truncation and its asymptotic dynamics}\label{sec:triad_truncation}

We introduce the triad truncation of~(\ref{e:eom_fourier}) for $\xi(k)\ne 0$, extending the approach of~\citet{kraichnan1973helical} who considered the case $\xi(k)\equiv 0$ corresponding to the inertial range approximation. We adopt the notation and build on the results of~\citet{moffatt2014note}. 

\subsection{Truncation}
Triad truncation is the projection of the dynamics~(\ref{e:eom_fourier}) onto three Fourier modes $\{\hat{\bs v}(\bs k,t),\hat{\bs v}(\bs p,t),\hat{\bs v}(\bs q,t)\}$ such that $\bs k+\bs p+\bs q=\bs 0$. The truncation is a first step beyond full linearization (which completely decouples the Fourier modes), to keep the smallest non-trivial portion of the quadratic nonlinearity. The velocity field reduces to
\be
\label{e:vfield_truncated}
\bs v(\bs x,t)=\hat{\bs v}(\bs k,t)e^{i\bs k \cdot \bs x}+\hat{\bs v}(\bs p,t)e^{i\bs p \cdot \bs x}+\hat{\bs v}(\bs q,t)e^{i\bs q \cdot \bs x}+\tn{c.c.},
\ee
where c.c. denotes complex conjugate terms which ensure that $\bs v(\bs x,t)$ is real. Since the triad $\{\bs k,\bs p, \bs q\}$ forms a triangle, it may be taken to lie in the $(x,y)$-plane by rotating the coordinate system, implying that the velocity field is independent of the spatial variable $z$. This allows one to introduce a stream function $\psi$ and write the velocity field as $\bs v=(\p \psi / \p y, -\p \psi/\p x, v_z)$. Thus, rather than working with the representation~(\ref{e:vfield_truncated}), it is more convenient to introduce the triadic expansions of the scalars $\psi$ and $v_z$~\citep{moffatt2014note}
\bse
\label{e:Moffatt_representation}
\be
\psi(x,y,t)&=&A_k(t)e^{i\bs k \cdot \bs x}+A_p(t)e^{i\bs p \cdot \bs x}+A_q(t)e^{i\bs q \cdot \bs x}+\tn{c.c.},  \\
v_z(x,y,t)&=&B_k(t)e^{i\bs k \cdot \bs x}+B_p(t)e^{i\bs p \cdot \bs x}+B_q(t)e^{i\bs q \cdot \bs x}+\tn{c.c.}.
\ee
\ese
Following step by step the derivation in~\citep{moffatt2014note}, the triad truncation of~(\ref{e:eom_fourier}) in terms of the complex vectors $\bs A=(A_k,A_p,A_q)$ and $\bs B=(B_k,B_p,B_q)$ results in the following system of coupled differential equations
\bse
\label{e:triad_dynamical_system}
\be
\label{e:Aeom}
 \mathsfbi I\bs{\dot A}+ \mathsfbi D  \mathsfbi I\bs A&=&2\Delta ( \mathsfbi I\bs A^* \times \bs A^*), \\
\label{e:Beom}
\bs{\dot B}+ \mathsfbi D\bs B&=&2\Delta (\bs B^* \times \bs A^*),
\ee
\ese
where $\Delta=(k_xp_y-k_yp_x)/2$ is the area of the triangle formed by $\{\bs k,\bs p, \bs q\}$ and
\be
 \mathsfbi I=\tn{diag}(k^2,p^2,q^2), \quad  \mathsfbi D=\tn{diag}(\xi(k),\xi(p),\xi(q)).
\ee
The positive and negative entries of $\mathsfbi D$ represent dissipation and forcing of the three modes, respectively. The key difference between the system~(\ref{e:triad_dynamical_system}) and the classical triad truncation is the matrix $\mathsfbi D$, which vanishes in the latter case. The typically studied case $\mathsfbi D=\mathsfbi 0$ is suitable for the inertial range considerations in classical turbulence and arises formally from the truncation of the inviscid Euler equation. In the context of active turbulence, we are interested in the case $\mathsfbi D\ne\mathsfbi 0$.
\par
 Energy $E$ and helicity $H$ of the triad are given by~\citep{moffatt2014note}
\bse
\be
2E&=&k^2|A_k|^2+p^2|A_p|^2+q^2|A_q|^2+|\bs B|^2, \\
H&=&\mathsfbi I\bs A \cdot \bs B^*+\mathsfbi I\bs A^* \cdot \bs B.
\label{e:helicity_triad}
\ee
\ese
In the remainder, we restrict our analysis to the triads obeying
\be
\label{e:pos_trace}
\tn{tr}(\mathsfbi D)=\xi(k)+\xi(p)+\xi(q)>0.
\ee
 Since in a finite spatial domain  the number of active modes  with $\xi(k)<0$ is finite, this condition is always satisfied for triads with at most two active legs, say $\xi(p)<0$ and $\xi(k)<0$ but $\xi(q)>0$, provided the forcing is sufficiently weak.

Finally, we express the helical decomposition~\citep{constantin1988beltrami,waleffe1992,alexakis2017helically} in terms of $\bs A$ and $\bs B$. Since the triad lies in the $(x,y)$-plane, the curl eigenmodes can be taken as
\be
\bs h_{\pm}(\bs k)&=&\hat{\bs z}\times \hat{\bs k}\pm i \hat{\bs z}=(-k_y,k_x,\pm i k)/k.
\ee
Projecting $\hat{\bs v}(\bs k)$ onto these eigenmodes gives the helical decomposition
\be
\label{e:helicity_decomp}
a_\pm(\bs k)=\f{1}{2}\bs h^\pm(\bs k)^*\cdot\hat{\bs v}(\bs k)=-\f{i}{2}(kA_k\pm B_k).
\ee
Analogous expressions hold for $\bs p$ and $\bs q$.

\subsection{Asymptotic rigid body dynamics: A cubic invariant}\label{sec:cubic_inv}
Since, according to~(\ref{e:triad_dynamical_system}), the dynamics of $\bs A$ affects $\bs B$, but not vice versa, we study Eqs.~(\ref{e:Aeom}) first. In components,~(\ref{e:Aeom}) reads
\be
\left. \begin{array}{ll}  
  k^2\dot{A}_k+\xi(k)k^2A_k=2\Delta(p^2-q^2)A_p^*A_q^*\\[8pt]
p^2\dot{A}_p+\xi(p)p^2A_p=2\Delta(q^2-k^2)A_q^*A_k^* \\[8pt]
q^2\dot{A}_q+\xi(q)q^2A_q=2\Delta(k^2-p^2)A_k^*A_p^*
 \end{array}\right\}.
  \label{e:Aeom_comp}
\ee
This system has the following three properties:
\settasks{	counter-format=(tsk[r]), label-width=4ex}
\begin{tasks}
\task 
If the initial conditions are real, then $\bs A(t)$ is real for all $t$. In this case, equations~(\ref{e:Aeom}) reduce to the Euler equations for the rotation of a rigid body.
\task 
The change of variables given by the constant phase shifts $(\phi_k,\phi_p,\phi_q)$
\be
(A'_k,A'_p,A'_q)=(A_k e^{-i\phi_k},A_p e^{-i\phi_p},A_q e^{-i\phi_q}) \;\; \tn{where } \phi_k+\phi_p+\phi_q=0
\qquad
\ee
leaves the equations~(\ref{e:Aeom_comp}) unchanged.
\task 
The following identity holds
		\be
		\Imag(A_kA_pA_q)=|A_k| |A_p| |A_q| \sin(\phi_k+\phi_p+\phi_q)=C\,\tn{exp}[-\tn{tr}(\mathsfbi D)t],
		\label{eq:property_iii}
		\ee
		where $C=\Imag[A_k(0)A_p(0)A_q(0)]$ and we introduced polar representations $A_k=|A_k| e^{i\phi_k}$, etc. Equation~(\ref{eq:property_iii}) also implies that
		\be
		\label{e:prop_iii_consequence}
		\Imag(k^2\dot{A}_k^*A_k)=
		k^2\det
		    \left[
      \begin{array}{cc}
     \Real\dot{A}_k  & \Real A_k         \\[0.3em]
     \Imag\dot{A}_k  & \Imag A_k
     \end{array}
    \right]     
=2\Delta(p^2-q^2)C\,\tn{exp}[-\tn{tr}(\mathsfbi D)t],
\quad
\ee
where we introduced the real and imaginary components, $A_k=\Real A_k+i \Imag A_k$. Analogous expressions hold for $A_p$ and $A_q$. Equation~(\ref{e:prop_iii_consequence}) has a useful geometrical interpretation: It gives the areal velocity (rate at which area is swept out) as a function of time of the complex trajectory traced out by the mode $A_k(t)$. Since we focus on triads with $\tn{tr}(\mathsfbi D)>0$, this immediately implies that the mode eventually vanishes, becomes stationary, or its trajectory approaches a line through the origin.
\end{tasks}
The property (i) was pointed out in~\citep{waleffe1992,moffatt2014note}. The second property is easily verified by direct substitution. To derive the last property, multiply the first equation in~(\ref{e:Aeom_comp}) by $A_pA_q$, etc., to obtain
\be
\left. \begin{array}{ll}  
  k^2\dot{A}_kA_pA_q+\xi(k)k^2A_kA_pA_q=2\Delta(p^2-q^2)|A_p|^2|A_q|^2\\[8pt]
p^2A_k\dot{A}_pA_q+\xi(p)p^2A_kA_pA_q=2\Delta(q^2-k^2)|A_q|^2|A_k|^2 \\[8pt]
q^2A_kA_p\dot{A}_q+\xi(q)q^2A_kA_pA_q=2\Delta(k^2-p^2)|A_k|^2|A_p|^2
 \end{array}\right\}.
\ee
Subtract from each equation its complex conjugate and add the resulting expressions
\be
\dot{A}_kA_pA_q+A_k\dot{A}_pA_q+A_kA_p\dot{A}_q+[\xi(k)+\xi(p)+\xi(q)]A_kA_pA_q-\tn{c.c}=0.
\ee
Now use the chain rule and substitute $\xi(k)+\xi(p)+\xi(q)=\tn{tr}(\mathsfbi D)$
\be
\f{d}{dt}\big(A_kA_pA_q-A_k^*A_p^*A_q^*\big)=-\tn{tr}(\mathsfbi D)\big(A_kA_pA_q-A_k^*A_p^*A_q^*\big).
\ee
Property (iii) then follows from integrating this first order equation. To derive~(\ref{e:prop_iii_consequence}), multiply the first equation in~(\ref{e:Aeom_comp}) by $A_k^*$, etc., subtract from each such obtained equation its complex conjugate and then use~(\ref{eq:property_iii}). 

We note that (iii) also implies that $\Imag(A_kA_pA_q)$ is conserved in the inertial range of classical turbulence, where $\mathsfbi D=\mathsfbi 0$ holds. This adds a cubic invariant to a list of quadratic invariants of the classical triadic system~\citep{waleffe1992,moffatt2014note,RathmannDitlevsen2017}. In section~\ref{sec:classical_triads_classification} we combine the cubic invariant with the conservation of in-plane energy and enstrophy~\citep{moffatt2014note} to obtain a detailed geometric classification of  the solutions of the system~(\ref{e:Aeom}) when $\mathsfbi D=\mathsfbi 0$.

\subsection{Asymptotic dynamics: rigid body and particle in a magnetic field}\label{subsec:asymptotic_dynamics}
We use the properties (i-iii) to argue that the dynamics~(\ref{e:Aeom_comp}) is asymptotically equivalent to that of a forced rigid body with principal moments of inertia $(k^2,p^2,q^2)$. Since we consider triads for which $\tn{tr}(\mathsfbi D)>0$, equation~(\ref{eq:property_iii})  suggests that the phase curves of~(\ref{e:Aeom_comp}) approach the following algebraic subset $S$ at an exponential rate
\be
\label{e:Aeom_attractor}
\Imag(A_kA_pA_q)=|A_k| |A_p| |A_q|  \sin(\phi_k+\phi_p+\phi_q)=0.
\ee
For the purposes of asymptotic analysis, we assume it is sufficient to consider initial conditions $\bs A(0)$ lying on the attractor $S$. There are two possibilities: 
\be
\label{e:three_possibilities}
|A_i|=0\;\tn{for some } i\in\{k,p,q\} \quad \tn{or} \quad \phi_k+\phi_p+\phi_q=n\pi.
\ee
Regardless which of the three conditions $\bs A(0)$ satisfies, the property (ii) implies it is always possible to perform a change of variables that makes $\bs A(0)$ a real vector without altering the dynamics~(\ref{e:Aeom_comp}). But then it follows from property (i) that $\bs A(t)$ is real for all~$t$. It is therefore plausible that the asymptotic dynamics of the system~(\ref{e:Aeom_comp}) is  equivalent to the asymptotic dynamics of the system
\be
\label{e:eom_rigidbody}
\mathsfbi I\dot{\bs\omega}+\mathsfbi{DI}\bs \omega&=& \mathsfbi I\bs \omega \times \bs \omega,
\ee
where $\bs \omega=(\omega_k,\omega_p,\omega_q)$ is a real vector. Equation~(\ref{e:eom_rigidbody}) has the structure of the Euler equations for a forced rigid body with inertia tensor $\mathsfbi I$ and angular velocity $\bs \omega$. When a triadic leg lies in the active or passive range, the rigid body is either forced or damped along the corresponding axis of inertia. Importantly, the forcing/damping is proportional to the component of angular momentum $\mathsfbi I\bs \omega$ along that axis. The system~(\ref{e:eom_rigidbody}) admits exact solutions corresponding to exponential growth or decay of rotations about one principal axis only, for example $\bs \omega=c(e^{-D_{kk} t},0,0)$.

We now focus on the asymptotic dynamics of the system for $\bs B$~(\ref{e:Beom}). Since by the above analysis $\bs A$ can be eventually taken to be the real vector $\bs \omega$, the real  and imaginary parts of $\bs B$ asymptotically decouple into two equations
\bse
\be
\Real\bs{\dot B}+\mathsfbi D\Real\bs B&=&\Real\bs B \times \bs \omega, \\
\Imag\bs{\dot B}+\mathsfbi D\Imag\bs B&=&-\Imag\bs B \times \bs \omega.
\ee
\ese
The first equation has the structure of Newton's equations for a forced particle with velocity $\bs u=\Real\bs B$ and charge $+1$  moving in a magnetic field~$\bs \omega$. The second equation describes an analogous dynamics with velocity $\Imag\bs B$ and charge $-1$. Since for real-valued $\bs \omega$ the helicity~(\ref{e:helicity_triad}) is determined by the real part of $\bs B$, we may conclude that the triadic system~(\ref{e:triad_dynamical_system}), in the long-time limit, becomes equivalent to the following equations for the real vectors $\bs \omega$ and $\bs u$
\bse
\label{e:triad_system_asymptotically}
\be
\label{e:triad_system_asymptotically_1}
\mathsfbi I\dot{\bs\omega}+\mathsfbi D\mathsfbi I\bs \omega&=& \mathsfbi I\bs \omega \times \bs \omega, \\
\bs{\dot u}+\mathsfbi D\bs u&=&\bs u \times \bs \omega.
\label{e:triad_system_asymptotically_2}
\ee
\ese
The second equation means that the angular velocity $\bs \omega$ of the forced rigid body acts as a magnetic field for a forced particle moving with velocity $\bs u$. In this notation, the triad helicity is the dot product between the rigid body angular momentum and the particle velocity
\be
H=2\mathsfbi I\bs\omega\cdot \bs u.
\ee
Thus, the helicity is positive when the particle moves in the direction of the angular momentum and negative when it moves in the opposite direction.

\section{Triad classification}
\label{sec:triad_classification}
We would like  to classify active triads according to their long-time behaviour. To this end, it is useful to develop first an intuitive understanding based on the asymptotic correspondence with the \lq rigid body and a particle in a magnetic field\rq{} system~(\ref{e:triad_system_asymptotically}). Subsequently, we will confirm the intuitive picture through explicit numerical simulations.

Without forcing, $\mathsfbi D=\mathsfbi0$ in~(\ref{e:triad_system_asymptotically_1}), the rigid body dynamics admits three fixed points, which correspond to constant angular velocity rotation about one of the three principal axes. Rotation about the small $(p^2)$ and large $(q^2)$ axes is stable, while rotation about the middle axis $(k^2)$ is unstable~\citep{arnold2013mathematical}. With forcing, $\mathsfbi D\ne \mathsfbi 0$, the linear part of~(\ref{e:triad_system_asymptotically_1}) promotes exponential growth of the mode for which $D_{ii}<0$ and damping of the remaining modes. It is conceivable that, when combined with the Eulerian nonlinearity, the coupled dynamical system~(\ref{e:triad_system_asymptotically}), and hence the system~(\ref{e:triad_dynamical_system}), becomes unstable when the rigid body is forced at the small or large principal axis, for in this case the nonlinearity does not counteract the exponential growth. However, when forced at the middle principal axis, the nonlinearity should induce motion about the remaining axes. Since these axes are dissipative, the system should soon realign with the middle principal axis, until the nonlinearity becomes dominant again, and so on. Numerical investigations presented in section~\ref{sec:sub_stable_triads} suggest that the dynamics (\ref{e:Aeom}) indeed approaches a limit cycle, although we do not rule out the possibility of more complicated attractors for some particular triads and parameters $D_{ii}$.

In all numerical simulations of~(\ref{e:triad_dynamical_system}) we use the polynomial dispersion relation $\xi(k)$ given by~(\ref{e:dispersion_relation_higher_order}) with parameters $(\Gamma_0,\Gamma_2,\Gamma_4)$ corresponding to the characteristic triple  ($\Lambda=75\,\mu$m, $\tau=6.4$s, $\kappa=8.4$ mm$^{-1}$), as studied in~\citep{2017SlomkaDunkel}. For time-stepping, we use the classical Runge--Kutta method (RK4).

\begin{figure}
  \centerline{\includegraphics[width=0.85\textwidth]{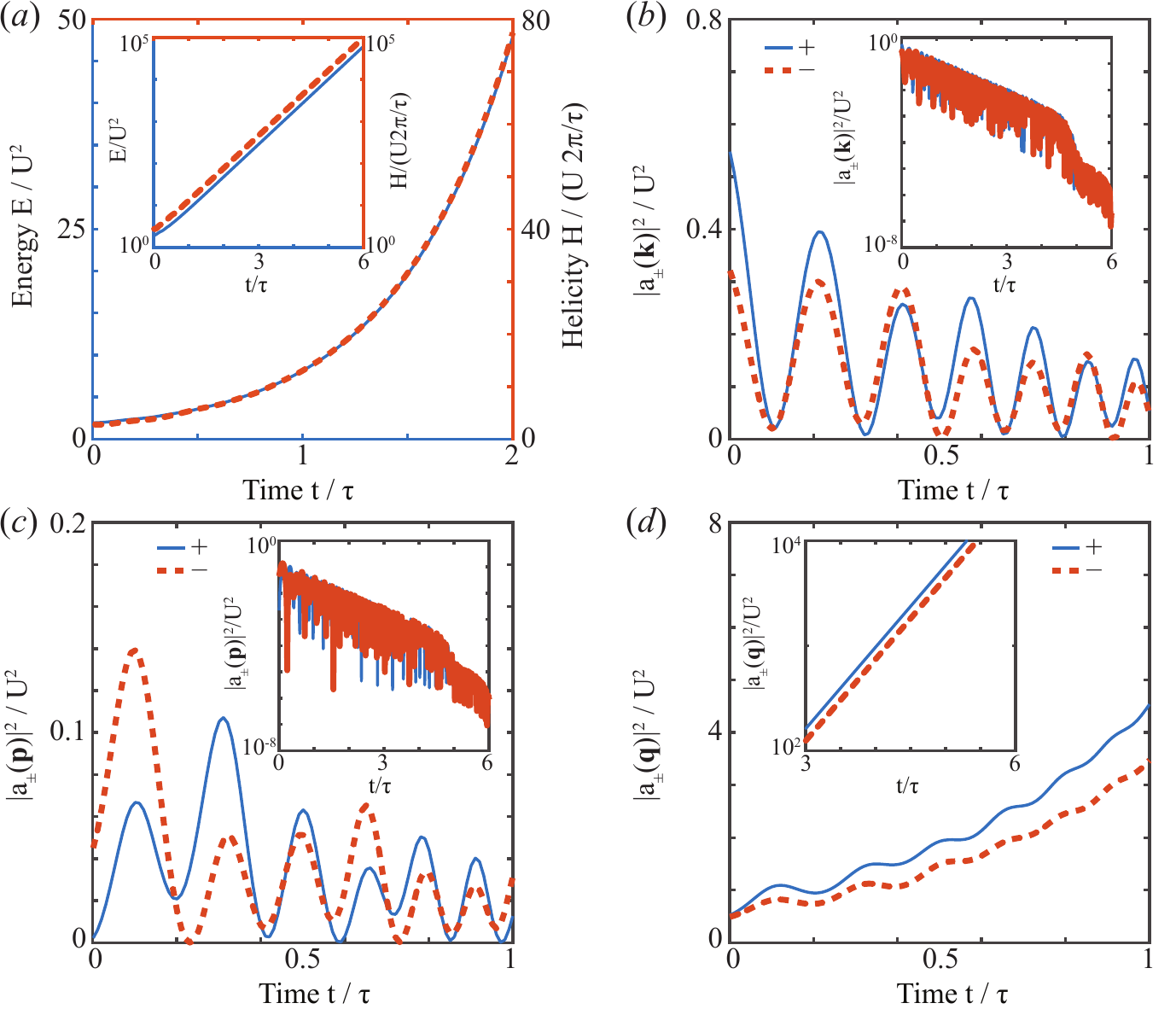}}
  \caption{Numerical simulations of~(\ref{e:triad_dynamical_system}) with polynomial dispersion~(\ref{e:dispersion_relation_higher_order}) initiated with random complex initial conditions show that active triads ($p<k<q$) are unstable when forced at large wavenumbers $q$. Energy and helicity increase exponentially (\textit{a}), reflecting the exponential growth of the forced helical mode (\textit{d}) and underdamped decay of the passive  helical modes (\textit{b, c}). Parameters: $\{\bs k,\bs p, \bs q\}=[(-5,9,0),(1,2,0),(4,-11,0)]$, box size $L=24\Lambda$.
  }
\label{fig:unstable_triads_kp_in_I}
\end{figure}

\begin{figure}
  \centerline{\includegraphics[width=0.85\textwidth]{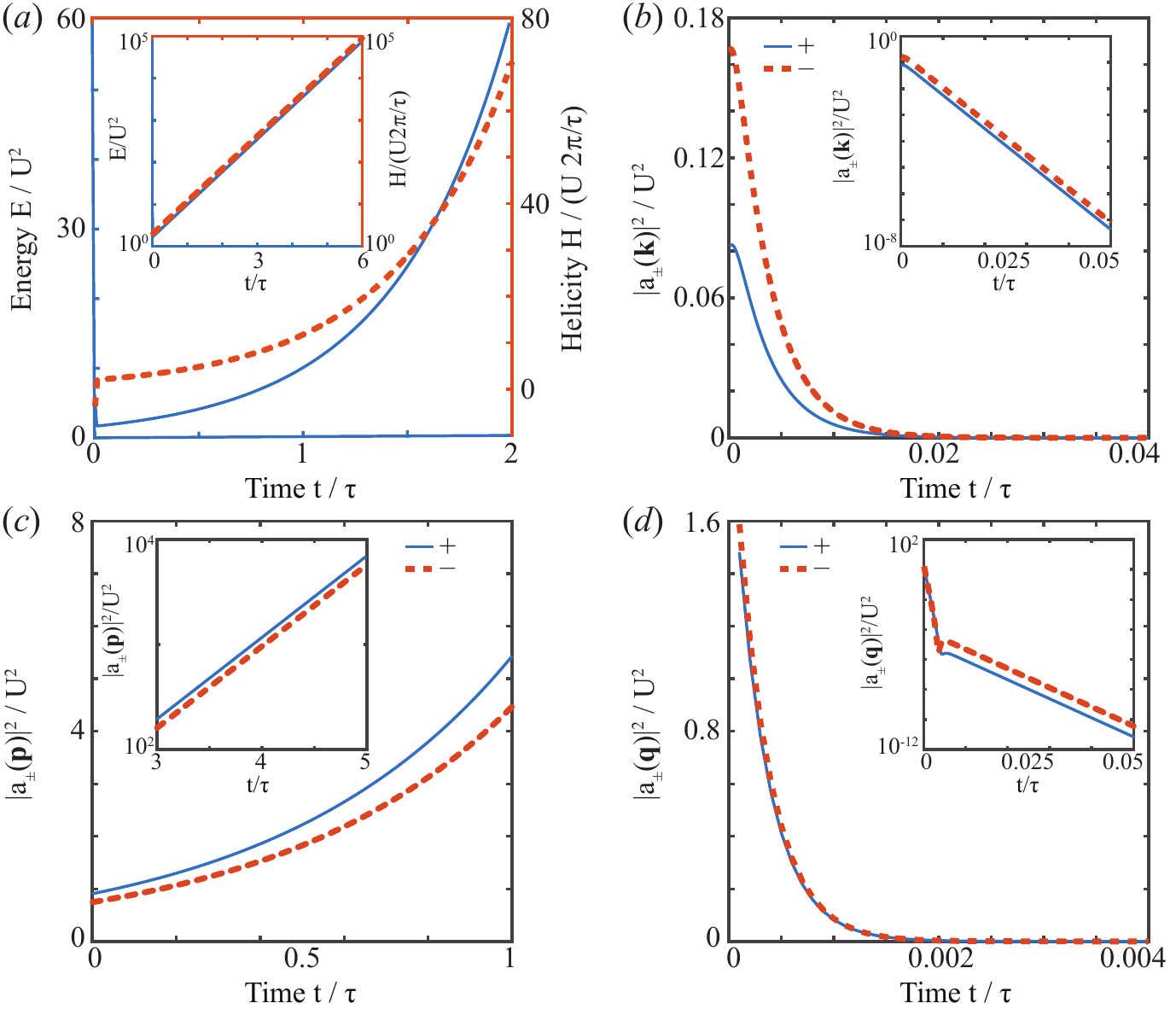}}
  \caption{
  Numerical simulations of~(\ref{e:triad_dynamical_system}) with polynomial dispersion~(\ref{e:dispersion_relation_higher_order}) initiated with random complex initial conditions  show that active triads ($p<k<q$) are unstable when forced at small wavenumbers $p$. Energy and helicity increase exponentially (\textit{a}), reflecting the exponential growth of the forced helical mode (\textit{c}) and overdamped decay of the passive helical modes (\textit{b, d}). Parameters: $\{\bs k,\bs p, \bs q\}=[(-14,-13,0),(4,-11,0),(10,24,0)]$, box size $L=24\Lambda$.
  }
\label{fig:unstable_triads_kq_in_III}
\end{figure}

\subsection{Unstable triads: rigid body forced at the small or large principal axis}\label{sec:unstable_triads}

Suppose the triadic system is forced at the small scale $q$, implying that $D_{qq}<0$ but $D_{kk}>0$ and $D_{pp}>0$ in~(\ref{e:triad_dynamical_system}). The rigid body correspondence suggests the $q$-mode should become unstable as the exponential  growth and the nonlinearity reinforce each other. Indeed, $\bs A=c(0,0,e^{-D_{qq} t})$ is an exact unstable solution of~(\ref{e:Aeom}). The remaining part of the triadic system~(\ref{e:triad_dynamical_system}) is the equation~(\ref{e:Beom}) for $\bs B$. In the long-time limit, when $\bs A\to c(0,0,e^{-D_{qq} t})$, we find the exact solution $\bs B=c'(0,0,e^{-D_{qq} t})$. Our numerical simulations suggest that this solution is an attracting phase curve for generic initial conditions, confirming the rigid body correspondence in this case, see figure~\ref{fig:unstable_triads_kp_in_I}.

The asymptotic growth of the forced modes $A_{q}$ and $B_{q}$ implies that both energy and helicity increase exponentially, as confirmed  in~figure~\ref{fig:unstable_triads_kp_in_I}(\textit{a}). Thus, at the level of a single triad, the mirror symmetry breaking may be generated by the following process in the full model~(\ref{e:eom_fourier}): the rigid body quickly approaches a state in which it is rotating about the $q$-axis, with the angular speed growing exponentially, while the particle accelerates in the direction of $q$ or in the direction directly opposite, producing positive or negative helicity, respectively, depending on initial conditions.  

A similar description characterised by exponential growth of energy and helicity applies when active triads are forced at the large scale $p$, see figure~\ref{fig:unstable_triads_kq_in_III}. What distinguishes the two types of forcing is the nature of the damping of the dissipative modes. When forced at large wavenumbers $q$, the decay is underdamped exhibiting oscillations, figure~\ref{fig:unstable_triads_kp_in_I}(\textit{b,c}), whereas forcing at the small wavenumbers $p$ results in overdamped dynamics, as shown in figure~\ref{fig:unstable_triads_kq_in_III}(\textit{b,d}), a direct consequence of the dependence of the damping force on the wavenumber magnitude. The asymptotic response of the system~(\ref{e:triad_dynamical_system}) when two modes are forced is identical to the above scenarios when one mode is forced, as discussed in the Appendix~\ref{app:triads_two_active_legs}.

\begin{figure}
  \centerline{\includegraphics[width=0.85\textwidth]{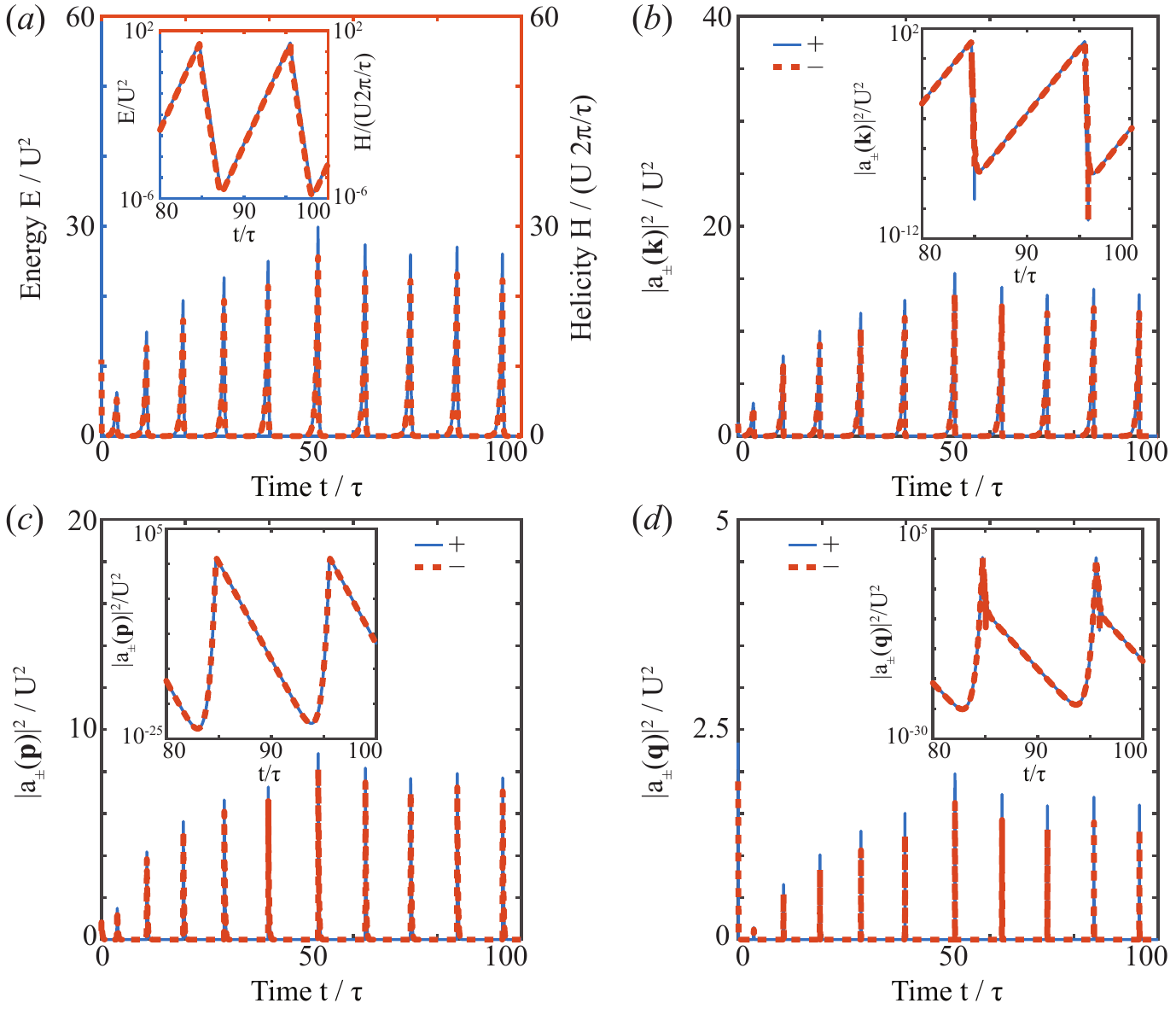}}
  \caption{
Numerical simulations of~(\ref{e:triad_dynamical_system}) with polynomial dispersion~(\ref{e:dispersion_relation_higher_order}) initiated with random complex initial conditions show that active triads ($p<k<q$) are stable when forced at intermediate scales $k$. The energy and helicity (\textit{a}) as well as the amplitudes of the helical modes (\textit{b--d}) stay bounded and soon take the form of very rapid charge-discharge bursts, reflecting the collapse of the dynamics onto a limit cycle, see figure~\ref{fig:limit_cycles}. Note the different $y$-scales in  (\textit{b--d}), which indicate that the energy produced by the intermediate scale is primarily send to large scales. This is a manifestation of the upward transfer at the level of a single triad. Parameters: $\{\bs k,\bs p, \bs q\}=[(12,1,0),(3,7,0),(-15,-8,0)]$, box size $L=24\Lambda$.}
\label{fig:stable_triads}
\end{figure}

\subsection{Stable triads: rigid body forced at the middle principal axis}\label{sec:sub_stable_triads}
For a rigid body forced at the middle principal axis we expect periodic behaviour since the nonlinearity destabilizes the action of the linear forcing in this case. Numerical simulations of~(\ref{e:triad_dynamical_system}) with  $D_{kk}<0$ but $D_{pp}>0$ and $D_{qq}>0$ show that the system equilibrates by developing periodic bursts characterised by alternating exponential growth and decay of energy, helicity and the helical modes, suggesting the existence of a stable limit cycle, see figure~\ref{fig:stable_triads}.

\begin{figure}
  \centerline{\includegraphics[width=1.0\textwidth]{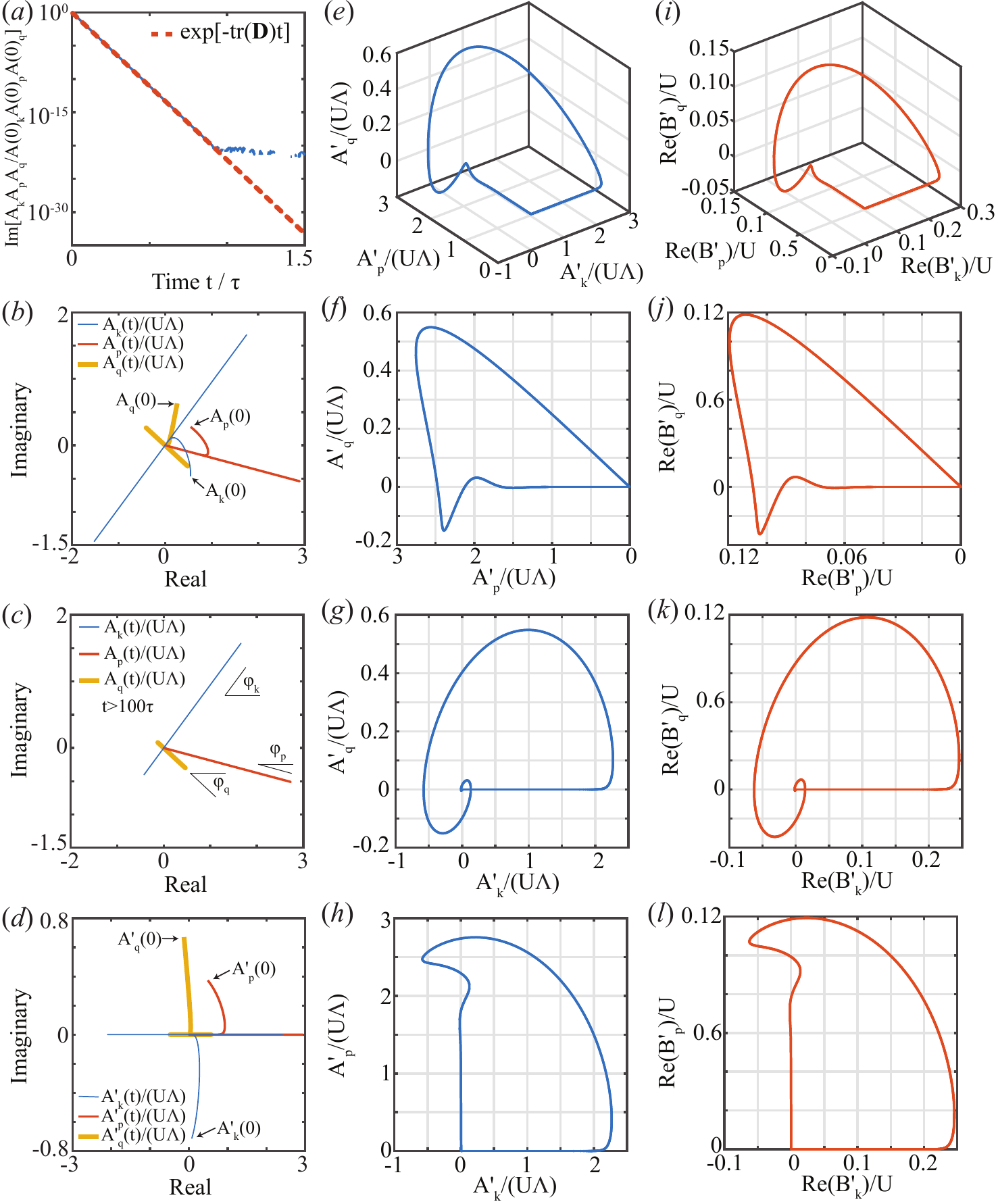}}
  \caption{
Asymptotic analysis of the results in~figure~\ref{fig:stable_triads}: the dynamics of stable active triads eventually collapses onto a limit cycle in a real three dimensional subspace.
(\textit{a}) The cubic quantity $\Imag (A_kA_pA_q)$ decays in accordance with~(\ref{eq:property_iii}) until the machine double-precision limit is reached.
(\textit{b}) Complex trajectories traced out by the modes $\bs A(t)=(A_k(t),A_p(t),A_q(t))$ approach straight lines at an exponential rate.
(\textit{c}) Trajectories in (\textit{b}) for $t>100\tau$. The lines are characterised by the angles $(\phi_k,\phi_p,\phi_q)=(0.759,-0.185,-0.574)$, such that $\phi_k+\phi_p+\phi_q=0$, as required by vanishing of $\Imag (A_kA_pA_q)$. 
(\textit{d}) The phases define the change of variables $ (A'_k,A'_p,A'_q)=(e^{-i\phi_k}A_k,e^{-i\phi_p}A_p,e^{-i\phi_q}A_q)$ and $ (B'_k,B'_p,B'_q)=(e^{-i\phi_k}B_k,e^{-i\phi_p}B_p,e^{-i\phi_q}B_q)$, which leaves the differential equations~(\ref{e:triad_dynamical_system}) unchanged, but rotates the complex trajectories so that the variables $(A'_k(t),A'_p(t),A'_q(t))$ become real in the limit $t\to\infty$.
(\textit{e}) In this three dimensional real subspace, $\bs A'(t)$ collapses onto a stable limit cycle.
(\textit{f--h}) Projections of the limit cycle of $\bs A'(t)$ onto the coordinate planes.
(\textit{i}) $\bs B'(t)$ also develops a limit cycle, shown is the real part.
(\textit{j--k}) The corresponding projections of $\Real\bs B'(t)$  onto the coordinate planes.
}
\label{fig:limit_cycles}
\end{figure}

To numerically verify the existence of a limit cycle in the system~(\ref{e:Aeom}) initiated with generic complex initial conditions, we now illustrate how to determine the three-dimensional real subspace onto which the system converges. We first note that the numerical solutions obey the property (iii) until the machine precision is reached; see figure~\ref{fig:limit_cycles}(\textit{a}). As a consequence, Eqs.~(\ref{e:prop_iii_consequence}) imply that each mode either vanishes, stops moving, or its trajectory in the complex plane approaches a line through the origin. In the present case of forcing the intermediate wavenumber $k$ all three modes follow the last scenario: the complex trajectories $(A_k(t),A_p(t),A_q(t))$ become straight lines, see figure~\ref{fig:limit_cycles}(\textit{b}),  with well-defined phase angles $(\phi_k,\phi_p,\phi_q)$, that satisfy $\phi_k+\phi_p+\phi_q=0$, see figure~\ref{fig:limit_cycles}(\textit{c}). We use these angles to define the change of variables $ (A'_k,A'_p,A'_q)=(e^{-i\phi_k}A_k,e^{-i\phi_p}A_p,e^{-i\phi_q}A_ q)$ and $ (B'_k,B'_p,B'_q)=(e^{-i\phi_k}B_k,e^{-i\phi_p}B_p,e^{-i\phi_q}B_q)$. This change of variables does not affect the Eqs.~(\ref{e:triad_dynamical_system}), it only rotates the complex trajectories so that the three modes $(A'_k,A'_p,A'_q)$ approach a real three-dimensional subspace at an exponential rate, figure~\ref{fig:limit_cycles}(\textit{d}). The asymptotic trajectory in that subspace reveals a limit cycle, figure~\ref{fig:limit_cycles}(\textit{e--h}), as expected from the rigid body correspondence. The limit cycle represents exponential growth of the rotation rate about the $k$-axis until the nonlinear effects destabilize it, followed by a rapid discharge along the two dissipative axes. The discharge along the $q$-axis represents energy transfer to small scales, while the discharge along the $p$-axis represents energy transfer to large scales. This behaviour likely explains, at the level of individual triadic interactions, the origin of the steady-state upscale energy transfer  in the full system~(\ref{e:eom_fourier}).

\subsection{Only stable triads admit a fixed point}
We still mention that the triadic system~(\ref{e:triad_dynamical_system}) forced at the intermediate wavenumber (and only in that case) exhibits a family of fixed points (see Appendix \ref{app:fixed_point} for details)
\bse
\label{e:fixed_point}
\be
\left[
\begin{array}{c}
     A_k        \\ 
     A_p       \\
     A_q
\end{array}     
\right]
     &=&
     \sqrt{\alpha}
\left[
\begin{array}{ccc}     
          \sqrt{|p^2-q^2|/|D_{kk}|}/k  \\[0.3em]
     \sqrt{|q^2-k^2|/|D_{pp}|}/p	\\[0.3em]
     \sqrt{|k^2-p^2|/|D_{qq}|}/q
\end{array}       
\right], 
\\  
\left[
\begin{array}{c}
     B_k           \\
     B_p 		\\
     B_q
\end{array}     
\right]   
     &=&  
\left[
\begin{array}{ccc}
     k \sqrt{|p^2-q^2|/|D_{kk}|}[c_1+ic_2(-k^2+p^2+q^2)]          \\[0.3em]
     p \sqrt{|q^2-k^2|/|D_{pp}|}[c_1+ic_2(k^2-p^2+q^2)]		\\[0.3em]
     q \sqrt{|k^2-p^2|/|D_{qq}|} [c_1+ic_2(k^2+p^2-q^2)]
\end{array}     
\right],
\ee
where
\be
\alpha=-\det(\mathsfbi I\mathsfbi D)/(4\Delta^2|p^2-q^2||q^2-k^2||k^2-p^2|).
\ee
\ese
The arbitrary real constants $c_1$ and $c_2$ determine energy and helicity. The property (ii) in section~\ref{sec:cubic_inv} also implies that we can rotate the solution in the complex plane provided the three phases sum to zero. The fixed points are unstable to linear perturbations. Notably, Eqs.~(\ref{e:fixed_point}) are also an exact stationary solution of the untruncated equations~(\ref{e:eom_fourier}).

\section{Implications for classical triads}\label{sec:classical_triads_classification}

In this section, we classify the geometry of the solutions of the equation~(\ref{e:Aeom}) for the case $\mathsfbi D= \mathsfbi 0$ corresponding to the triad truncation of the Euler equations
\be
\label{e:Aeom_Dzero}
 \mathsfbi I\bs{\dot A}&=&2\Delta ( \mathsfbi I\bs A^* \times \bs A^*).
\ee
The system~(\ref{e:Aeom_Dzero}) exhibits three constants of motion
\begin{eqnarray}
    \label{e:three_constants_MT}
    \left. \begin{array}{r}  
        k^2 |A_k|^2+p^2 |A_p|^2+q^2 |A_q|^2=E \\
        k^4 |A_k|^2+p^4 |A_p|^2+q^4 |A_q|^2=\Omega \\
        A_kA_pA_q-A_k^*A_p^*A_q^*=C
    \end{array}\right\}.
\end{eqnarray}
The quadratic constants $E$ and $\Omega$ were found by~\citet{moffatt2014note}, the new cubic constant $C$ was derived in section~\ref{sec:cubic_inv} above. The triple~(\ref{e:three_constants_MT}) suggests that the system~(\ref{e:Aeom_Dzero}) is confined to a three-dimensional surface in a six-dimensional phase space. We next summarize a series of results classifying the solutions to~(\ref{e:Aeom_Dzero}), which are rigorously proven in the Appendix~\ref{app:torus}.

\begin{figure}
  \centerline{\includegraphics[width=1.0\textwidth]{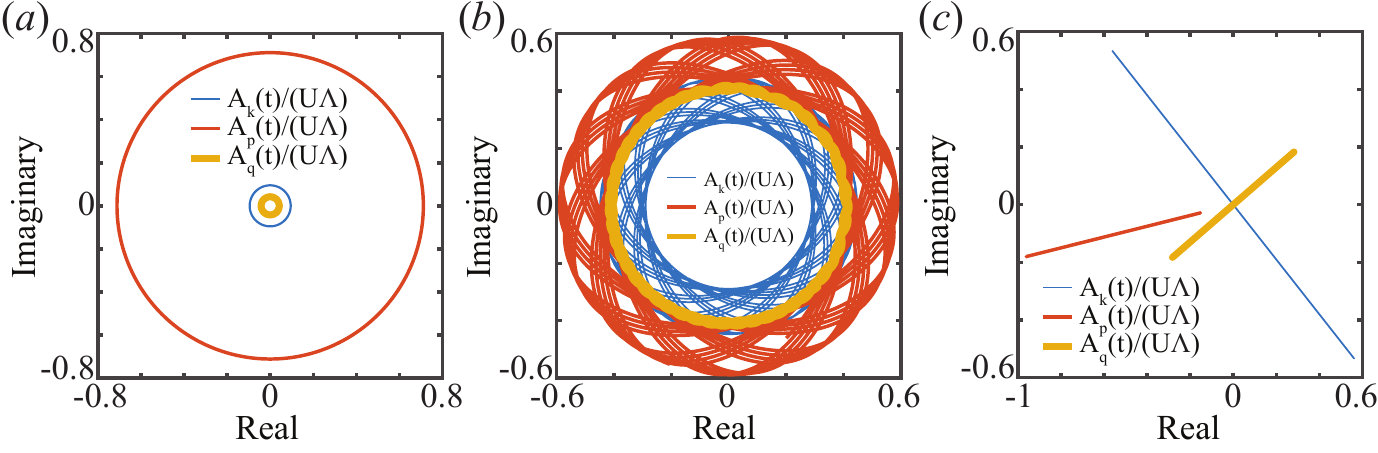}}
  \caption{
Types of orbits $\bs A(t)$ in the complex plane for the classical system~(\ref{e:Aeom})  with $\mathsfbi D=\mathsfbi0$ include: fixed points (not shown), circular orbits for initial conditions in the set $Z_3$~(\textit{a}), orbits resulting from trajectories on a three-torus for generic initial conditions~(\textit{b}), straight lines for initial conditions with $C=0$ in~(\ref{e:three_constants_MT}), in which case the system reduces to the classical Euler equations for a rigid body~(\textit{c}).
}
\label{fig:classical_triads}
\end{figure}

In the six-dimensional phase space for the system for $\bs A(t)$, we consider separately the following subsets of $\mathbb{R}^6$
\begin{eqnarray}
    \label{e:Zsubsets}
    Z_1&=&\{\, A_p=0, A_q=0\}\cup \{\, A_q=0,  A_k=0\} \cup \{\, A_k=0,  A_p=0\}, \\
        Z_3 &=& \{\,|A_k||A_p||A_q|\neq 0, 
        \,\mathrm{Re}(A_kA_pA_q) = 0\} \cap \nonumber\\
     && \{\, 
      |A_q|^2 |A_k|^2 k^2 q^2 (k^2 - q^2) 
    + |A_p|^2 |A_k|^2 p^2 k^2 (p^2 - k^2) 
    + |A_p|^2 |A_q|^2 q^2 p^2 (q^2 - p^2)
    = 0 \} \nonumber
\end{eqnarray}
Initial conditions in $Z_1$ correspond to fixed points of~(\ref{e:Aeom_Dzero}). For initial conditions in $Z_3$, the system~(\ref{e:Aeom_Dzero}) is solved exactly by a quasi-periodic motion with constant amplitudes and phases evolving linearly in time according to 
\begin{eqnarray}
    \left. \begin{array}{ll}  
        \phi_k = \pm(p^2-q^2) |A_p||A_q| / (k^2|A_k|) t + c_k\\[8pt]
        \phi_p = \pm(q^2-k^2) |A_q||A_k| / (p^2|A_p|) t + c_p\\[8pt]
        \phi_q = \pm(k^2-p^2) |A_k||A_p| / (q^2|A_q|) t + c_q
    \end{array} \right\},
\end{eqnarray}
where the equalities hold modulo $2\pi$ and the constants $c_i$ are chosen so that $\phi_k+\phi_p+\phi_q=\pi/2$ or $\phi_k+\phi_p+\phi_q=3/2\pi$ holds, as required by the definition of $Z_3$. Importantly, for initial conditions in $Z_3$ the sum of phases is conserved, so the system~(\ref{e:Aeom_Dzero}) stays in $Z_3$ and the phase space in fact can be reduced to a torus~$\mathbb{T}^2$. A typical trajectory for initial conditions on $Z_3$ is shown in figure~\ref{fig:classical_triads}(\textit{a}). In summary, fixed points and quasi-periodic motion completely characterise the solutions of~(\ref{e:Aeom_Dzero}) for initial conditions in $Z_1$ and $Z_3$, respectively.

We now consider the most important generic case of initial conditions in the complement $N=\mathbb{R}^6\backslash(Z_1\cap Z_3)$. In $N$, the differential of~(\ref{e:three_constants_MT}) has full rank, implying that~(\ref{e:three_constants_MT}) defines a three-dimensional manifold in $N$, that is, the solutions of~(\ref{e:Aeom_Dzero}) are confined to a smooth three-dimensional surface. For generic values of the triple $(E,\Omega,C)$, this surface is in fact a three-torus $\mathbb{T}^3$ (or several copies of such tori). A typical trajectory in such a generic case is shown in figure~\ref{fig:classical_triads}(\textit{b}). There are also special cases of $(E,\Omega,C)$ for which the manifold looks like (copies of) a product of a line and a torus $\mathbb{R}\times \mathbb{T}^2$ and/or (copies of)  $\mathbb{T}^3$. The reader is refered to Appendix~\ref{app:torus} for more details and rigorous proofs.
\par
Finally,  we still note that, since the solutions $\bs A(t)$ remain continuous and bounded for all~$t$, the linear system for $\bs B(t)$ can be solved exactly, at least formally, in terms of time-ordered matrix exponentials~\citep{gantmakher1998theory}.

\section{Gaussian active turbulence model}\label{sec:other_models}
The behaviour of individual active triads suggests that the mirror-symmetry breaking and upward energy transfer observed in the GNS system~(\ref{e:eom_fourier}) is first triggered by unstable active triads and then sustained by stable active triads. To test this hypothesis, we numerically study an alternative GNS model where the dispersion relation $\xi(k)$ in~(\ref{e:eom_fourier}) has the form
\be
\label{e:xi_newmodel}
\xi(k)=\Gamma_0k^2-\alpha\exp[-(k-k_0)^2/(2\sigma^2)].
\ee
The main difference between~(\ref{e:xi_newmodel}) and the polynomial model~(\ref{e:eom}) is that Gaussian activity model (\ref{e:xi_newmodel}) behaves like a Newtonian fluid with viscosity $\Gamma_0$ at both large and small scales, see~figure~\ref{fig:dispersion_relations}(\textit{a}). Equation~(\ref{e:xi_newmodel}) leads to an integro-partial differential equation in position space. In our simulations, we always fix $\Gamma_0=10^{-6}\,\tn{m}^2\tn{s}^{-1}$, corresponding to the kinematic viscosity of water. To relate the parameters $(\alpha, k_0, \sigma)$ to the characteristic triple $(\Lambda,\tau,\kappa)$, we must solve
\be
\label{e:Lambda_tau_kappa_definition}
\xi'(k_\Lambda)=0, \quad
\tau=-\xi^{-1}(k_\Lambda), \qquad
\xi(k_\pm)=0, \quad
\kappa=k_+-k_-,
\ee
where $k_\Lambda=\pi/\Lambda$ is the most unstable wavenumber and $k_\pm$ are the non-trivial zeros of the dispersion relation $\xi(k)$. Since no closed-form solutions exist, we solve the system~(\ref{e:Lambda_tau_kappa_definition}) numerically. We set $(\alpha, k_0, \sigma)=(2.544165\tn{ms}^{-1},\,52.36\tn{mm}^{-1},\,10\tn{mm}^{-1})$, yielding $(\Lambda,\tau,\kappa)=(65.14\mu\tn{m},\,0.1\tn{s},\,1.94\tn{mm}^{-1})$, which is in the range of typical bacterial suspension values~\citep{2017SlomkaDunkel}. Non-dimensionalising according to
\be
x= \f{L}{2\pi}\tilde x, \quad t= T\tilde t, \quad v= \f{L/(2\pi)}{T}\tilde v, \quad k= \f{2\pi}{L}\tilde k,
\ee
gives, after dropping the tildes and setting $T=(L/2\pi)^2/\Gamma_0$,
\be
\xi(k)=k^2-T\alpha\exp\big\{-[(2\pi/L)k-k_0]^2/(2\sigma^2)\big\}.
\ee
We simulate the dimensionless system in the vorticity-vector potential formulation as described in~\citep{2017SlomkaDunkel} using the Fourier pseudo-spectral method with the \lq3/2\rq-rule~\citep{SpecMethodsFD}, discretisation size $N=243^3$ and time step $dt=5\times 10^{-4}\tau/T$. We set the domain size $L=42\Lambda$, which corresponds to the most unstable wavenumber at $k_\Lambda=21$. For time-stepping, we use a third-order semi-implicit backward differentiation scheme~\citep{ascher1995implicit}.

\begin{figure}
  \centerline{\includegraphics[width=1.0\textwidth]{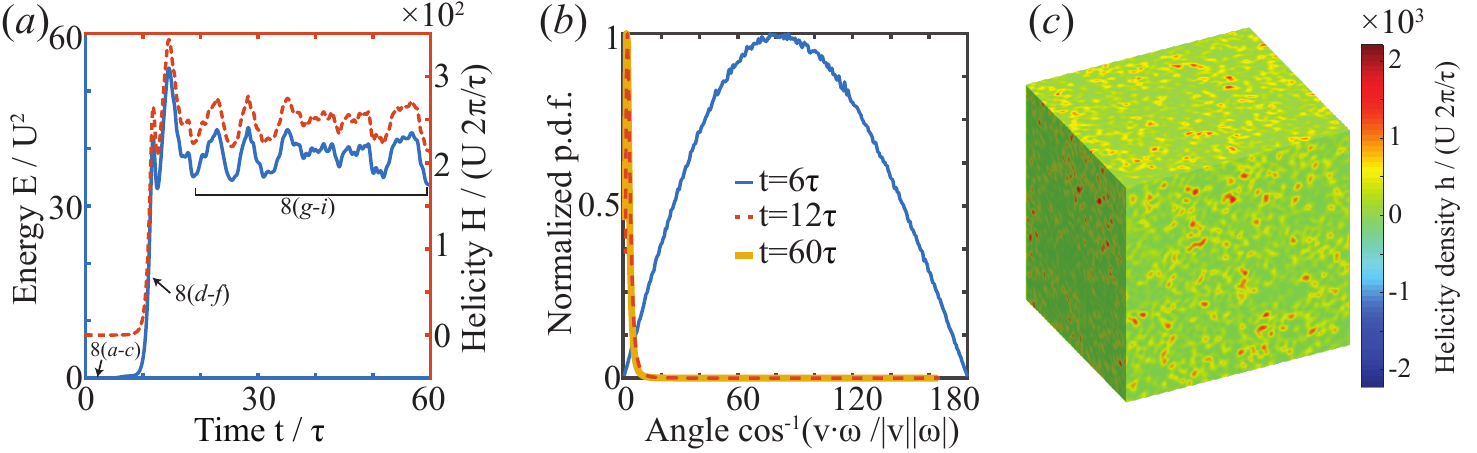}}
  \caption{
 Simulation results for the Gaussian model~(\ref{e:xi_newmodel}). 
(\textit{a})  Energy and helicity time series show the initial relaxation phase and the subsequent statistically stationary stage. Time instants and interval labels refer to figure~\ref{fig:spectrum_flux_triads}.
(\textit{b})  Normalized histograms of the angles between velocity~$\bs v$ and vorticity~$\bs \omega$ at three different time instants confirm that mirror-symmetry breaking is achieved by developing Beltrami-type flows, where velocity and vorticity are nearly aligned.
(\textit{c}) Snapshot of the helicity density field at $t=60\tau$ showing spontaneous symmetry breaking towards positive values.
}
\label{fig:xi_energy_helicity}
\end{figure}

\begin{figure}
  \centerline{\includegraphics[width=1.0\textwidth]{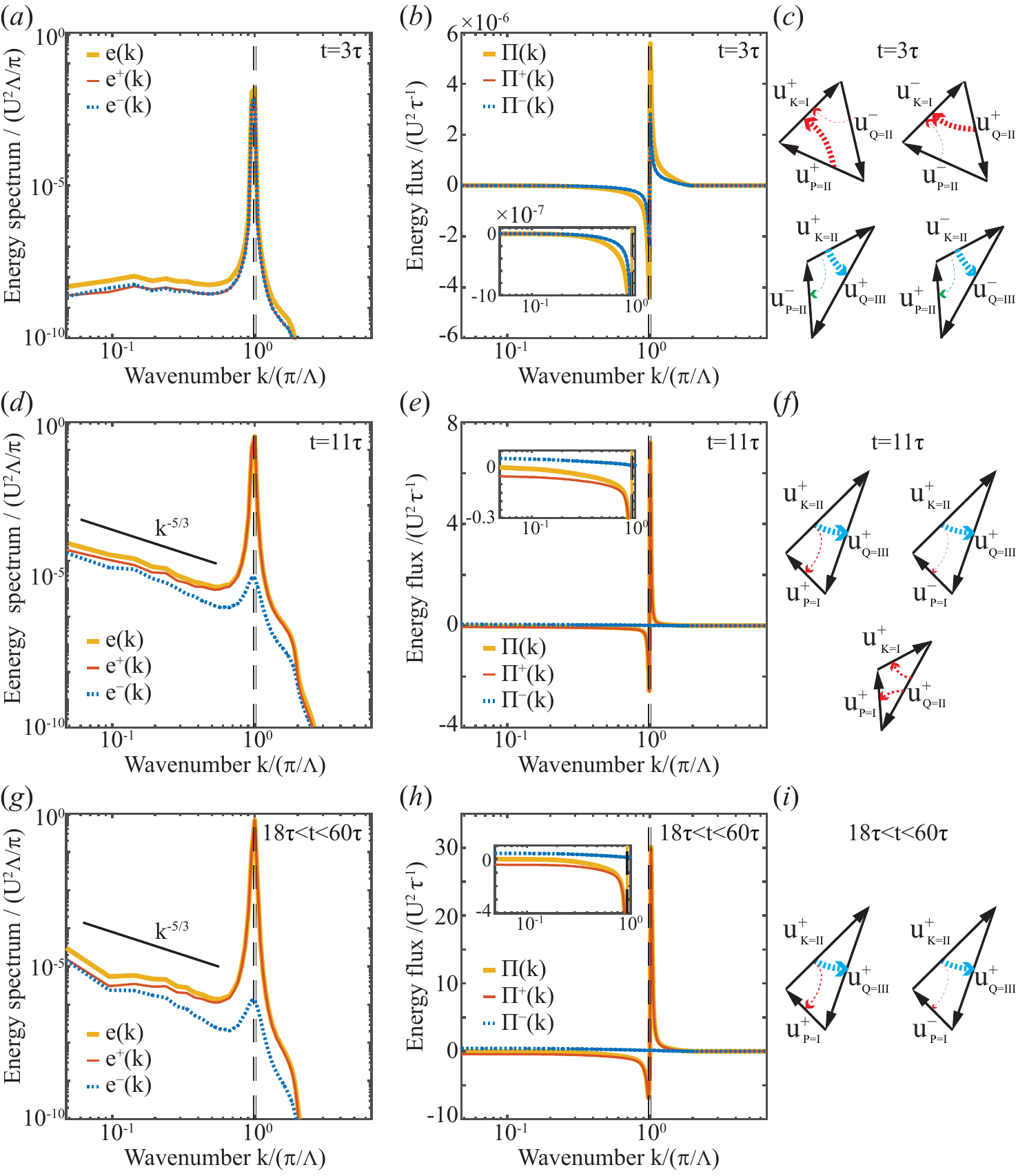}}
  \caption{
Numerical results for the Gaussian activity model~(\ref{e:xi_newmodel}) based on the simulation in figure~\ref{fig:xi_energy_helicity}.
Instantaneous~(\textit{a-f}) and average~(\textit{g-i}) energy spectra, fluxes and dominant integrated triads for time instants and intervals indicated in figure~\ref{fig:xi_energy_helicity}(\textit{a}). Vertical dashed lines mark the energy injection range.
}
\label{fig:spectrum_flux_triads}
\end{figure}
To discuss the results of numerical simulations, we use the helical decomposition~\citep{constantin1988beltrami,waleffe1992} to expand the velocity field in an orthogonal basis of curl operator eigenvectors $\bs h^\pm$
\be
\label{e:vel2hmodes}
\bs v (t,\bs k)=u^+(t,\bs k)\,\bs h^+(\bs k)+u^-(t,\bs k)\,\bs h^-(\bs k),
\ee
where $\bs h^\pm$  satisfy \mbox{$i\bs k \times \bs h^\pm=\pm k \bs h^\pm$} with $k=|\bs k|$. The decomposition~(\ref{e:vel2hmodes}) yields a splitting into cumulative energy and flux contributions  $e^\pm(k)$ and  $\Pi^\pm(k)$ from helical modes $u^\pm (\bs k)$ lying on the wavenumber shell $k$. Specifically, $\Pi^+(k)=\sum_{i=1}^4\Pi^i(k)$ and $\Pi^-(k)=\sum_{i=5}^8\Pi^i(k)$, where $\Pi^i(k)$ is one of the eight types of helicity-resolved fluxes and the summation follows the binary ordering of~\cite{waleffe1992}.  To analyse which triads are spontaneously activated at various time instants, we consider combinations  $K,P,Q\in\{\mathrm{I,II,III}\}$ of the three spectral domains in figure~\ref{fig:dispersion_relations}(\textit{b}), with region I corresponding to large scales, II to the energy injection range and III to small scales,  and distinguish modes by their helicity index  $s_K, s_P, s_Q\in\{\pm\}$. The helicity-resolved  integrated energy flow into the region $(K,s_K)$ due to interaction with regions $(P,s_P)$ and $(Q,s_Q)$ is given by
\be\label{e:symm_triad}
\mcal{T}_{KPQ}^{s_K s_P s_Q}=\f{1}{2}\big(\tilde{\mcal{T}}_{KPQ}^{s_K s_P s_Q}+\tilde{\mcal{T}}_{KQP}^{s_K s_Q s_P}\big),
\ee
where the unsymmetrized flows are defined by
\be
\label{e:regionhelicalint}
\tilde{\mcal{T}}_{KPQ}^{s_K s_P s_Q}=-\int d^3 x\, \bs v^{s_K}_K\cdot [(\bs v^{s_P}_P\cdot\nabla)\bs v^{s_Q}_Q], 
\ee
with $\bs v^{s_K}_K(t,\bs x)$ denoting the helical Littlewood-Paley velocity components, obtained by projecting on modes of a given helicity index $s_K\in\{\pm\}$ restricted to the Fourier space domain~$K$. Entries of the tensor $\mcal{T}$ are large when the corresponding triads are dominant. For example, a positive (negative) value of $\mcal{T}_{\mathrm{I,II,III}}^{+++}$ indicates that energy flows into (out of) large scale (I) positive helicity modes due to interactions of these modes with positive helical modes corresponding to energy injection range (II) and small scales (III).

Our numerical simulations show that the Gaussian-forcing model~(\ref{e:xi_newmodel}) and the polynomial model~(\ref{e:dispersion_relation_higher_order}) exhibit qualitatively similar behaviour, cf.~figures~\ref{fig:xi_energy_helicity},~\ref{fig:spectrum_flux_triads}  and corresponding plots in~\citep{2017SlomkaDunkel}. The Gaussian activity model also undergoes mirror symmetry breaking and spontaneously develops a non-zero net helicity, by realising chaotic Beltrami-type flow states in which velocity $\bs v$ and vorticity $\bs\omega$ are almost aligned, see figure~\ref{fig:xi_energy_helicity}. Figure~\ref{fig:spectrum_flux_triads} shows instantaneous and time-averaged energy spectra, energy fluxes and the dominant entries of the integrated triadic energy flows~(\ref{e:symm_triad}) for the time instants and intervals marked in figure~\ref{fig:xi_energy_helicity}(\textit{a}). The energy spectra in~figures~\ref{fig:spectrum_flux_triads}(\textit{a,d,g}) indicate that the system spontaneously selects positive helicity modes at all relevant wavenumbers in this particular realisation, while the energy fluxes  in~figures~\ref{fig:spectrum_flux_triads}(\textit{b,e,h}) are always negative at scales larger than the energy injection range (vertical dashed lines),  demonstrating the inverse energy cascade. Unlike the polynomial model, however, the long-time spectra of the Gaussian activity model develop an approximate Kolmogorov $-5/3$ scaling at large wavelengths, see figure~\ref{fig:spectrum_flux_triads}(\textit{d}). Note that, in the statistically stationary stage, the upward transfer is balanced by viscous dissipation; that is, no additional large-scale dissipation is required in the simulations. The dominant integrated energy flows shown in figures~\ref{fig:spectrum_flux_triads}(\textit{c,h,i}), where broken arrows indicate the direction of the inter-scale energy transfer and their thickness the relative magnitude of the transfer, are in agreement with the hypothesis that unstable triads drive the initial relaxation until stable triads become dominant and sustain the statistically stationary chaotic flow states.

\section{Conclusions}

We derived a previously unknown cubic invariant for the triad dynamics and used it to analyse and compare the triad truncations of two generalized Navier-Stokes (GNS) models and the classical Euler equations. In the GNS case, we focused on active triads with one or two modes in the energy injection range and found that their dynamics is asymptotically equivalent to a coupled system consisting of a forced rigid body and a forced particle in a magnetic field. This analogy allows one to distinguish unstable and stable active triads, based on whether the rigid body is forced along the small/large principal axes (large/small scales) or the middle principal axis (intermediate scales), respectively. The dynamics of the active GNS triads differs strongly from those of the classical Euler triads, for which the rigid body analogy does not hold in general and solutions are confined to a three-torus for generic initial conditions~(section~\ref{sec:classical_triads_classification}).

\par
The existence of unstable and stable triads explains recent  numerical results in~\citep{2017SlomkaDunkel}, which suggested that the polynomial 3D GNS models can spontaneously break mirror symmetry by developing Beltrami-like flow states and upward energy transfer: Unstable triads induce exponential helicity growth from small perturbations and dominate the initial relaxation. Because of the nonlinear coupling between the triads, the stable triads eventually become dominant and the system settles into a statistically stationary  chaotic flow state. In the stationary regime, energy is transferred from the spectral injection range to both large and small scales. This is consistent with the behaviour of stable triads, which develop a limit cycle. In the rigid body analogy, this limit cycle represents a periodic two-phase process. During the first phase, the rigid body  accumulates energy by increasing its spinning rate along the middle principal axis; during the second phase, the accumulated energy is released along the small and large principal axes. This release of the energy corresponds to energy transfer to large and small scales in the untruncated hydrodynamic equations. We confirmed the above picture for an alternative GNS model~(\ref{e:xi_newmodel}), which combines viscous dissipation and  active Gaussian forcing,  by computing the integrated energy flow between the three spectral domains (large scales, energy injection range and small scales). Unlike the previously studied polynomial model, the Gaussian active turbulence model develops energy spectra that approximately follow Kolmogorov's $-5/3$ scaling at large wavelengths, which may be desirable in applications to microbial suspensions.

\par
More broadly, the above results  suggest that parity violation and an inverse energy cascade may be generic features of turbulence models where the forcing term depends on the velocity field. The degree to which the mirror-symmetry is broken or the proportion of energy that is transferred to small and large scales should depend on the particular forcing considered. The two GNS models~(\ref{e:dispersion_relation_higher_order}) and~(\ref{e:xi_newmodel}) analysed here are basic examples that introduce a bandwidth of linearly unstable modes. These models can help guide theoretical efforts to find other forcing schemes that realize specific desired features, such as the magnitude of the upward transfer or its inertial character. Biological and engineered active fluids are promising candidates for the experimental implementation, as GNS models can be fitted to reproduce experimentally observed velocity correlation functions~\citep{2017SlomkaDunkel}. However, the general nature of the triad-based arguments presented here suggests that other non-equilibrium fluids might also be capable of breaking mirror-symmetry and developing upward energy transfer.  Last but not least, our results indicate that helical flows~\citep{moffatt2014helicity} and the Beltrami-type flows in particular, which have been primarily studied as exact stationary solutions of the Euler equations~\citep{arnold1999topological} and  in the context of magnetodynamics~\citep{marsh1996force,yoshida2001beltrami,hudson2007eigenvalue}, could be more ubiquitous than previously thought.

The authors thank Ruben Rosales, Luiz Faria and Vili Heinonen for helpful discussions. This work was supported by an Alfred P. Sloan Research Fellowship (J.D.), an Edmund F. Kelly Research Award (J.D.) and a Complex Systems Scholar Award of the James S. McDonnell Foundation (J.D.).



\appendix

\section{Triads forced at two legs}\label{app:triads_two_active_legs}
\begin{figure}
  \centerline{\includegraphics[width=0.9\textwidth]{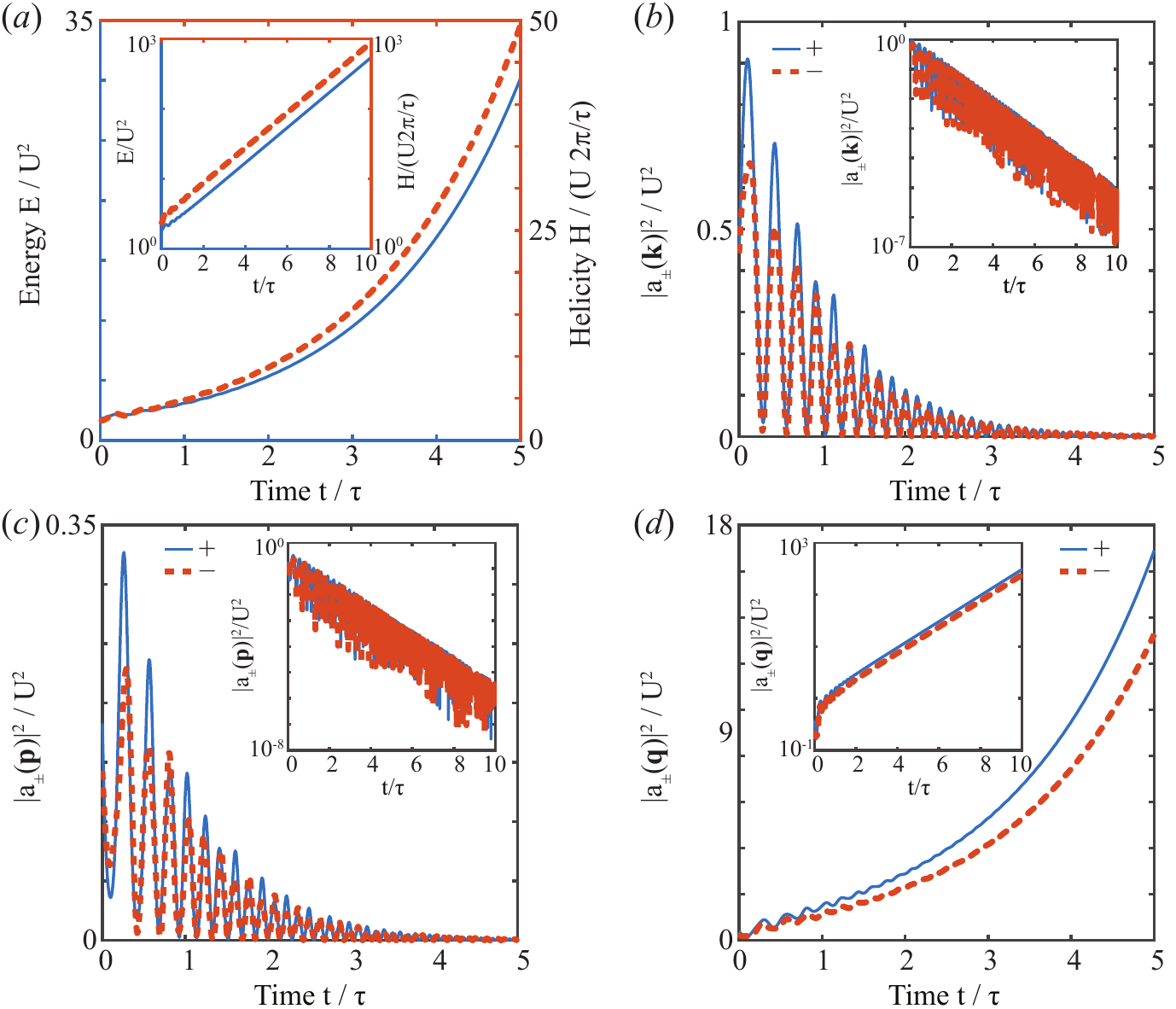}}
  \caption{
Numerical simulations of the triad dynamics~(\ref{e:triad_dynamical_system}) initiated with generic complex initial conditions show that active triads ($p<k<q$) are unstable when forced at intermediate $k$ and small $q$ scales. Energy and helicity increase exponentially~(\textit{a}), reflecting the exponential growth of one of the forced modes~(\textit{d})  and underpdamed decay of the remaining forced mode~(\textit{b})  and the passive mode~(\textit{c}). Parameters: $\{\bs k,\bs p, \bs q\}=[(4,-11,0),(-9,-1,0),(5,12,0)]$, box size $L=24\Lambda$.
}
\label{fig:app_unstable_p_in_I_kq_in_II}
\end{figure}

\begin{figure}
  \centerline{\includegraphics[width=0.9\textwidth]{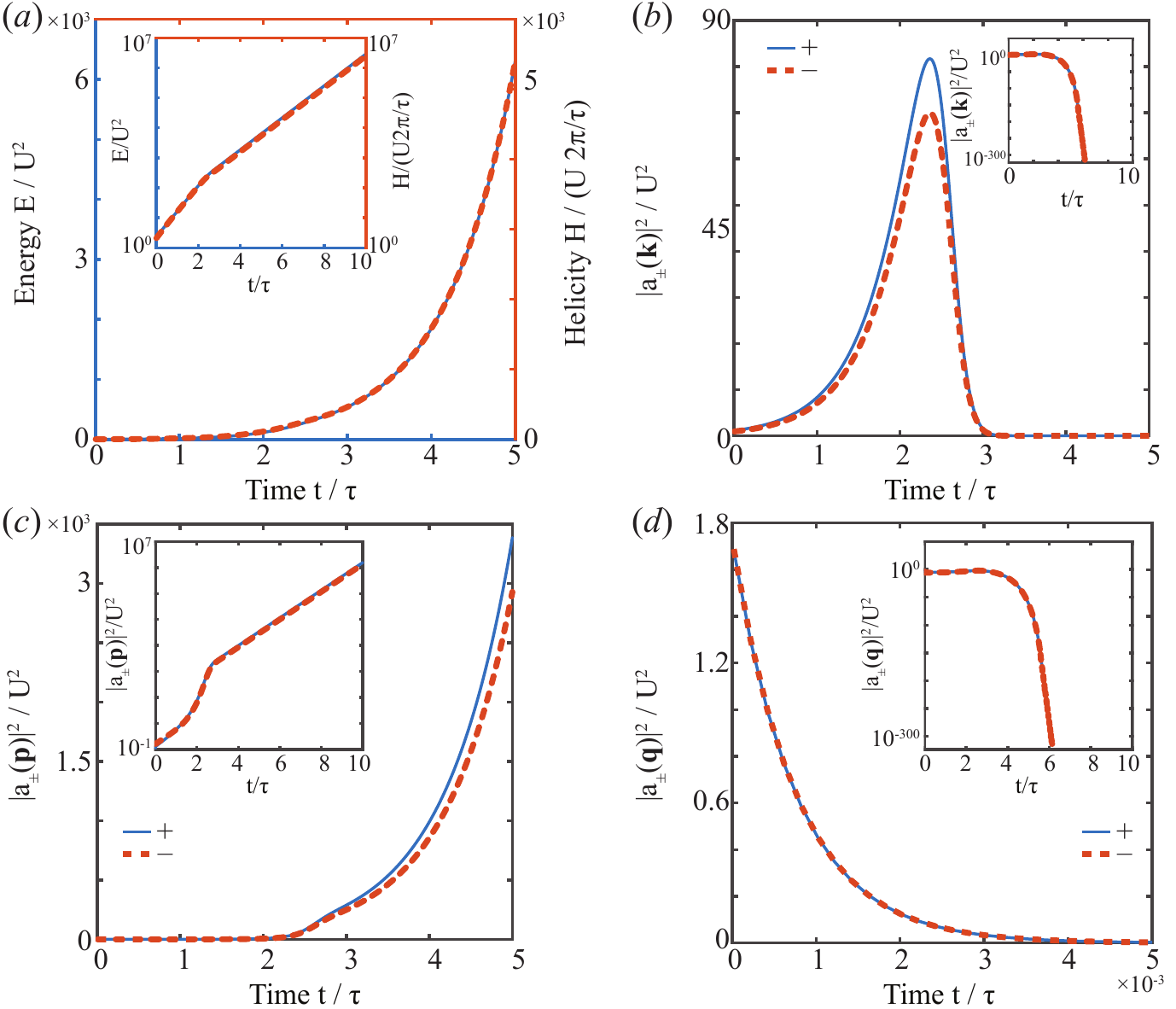}}
  \caption{
Numerical simulations of the triad dynamics~(\ref{e:triad_dynamical_system}) initiated with generic complex initial conditions show that active triads ($p<k<q$) are unstable when forced at large $p$ and intermediate $k$ scales. Energy and helicity increase exponentially~(\textit{a}), reflecting the exponential growth of one of the the forced modes~(\textit{c}) and overdamped decay of the remaining forced mode~(\textit{b})  and the passive mode~(\textit{d}). Parameters: $\{\bs k,\bs p, \bs q\}=[(5,11,0),(8,8,0),(-13,-19,0)]$, box size $L=24\Lambda$.
}
\label{fig:app_unstable_pk_in_II_q_in_III}
\end{figure}

Figures~\ref{fig:app_unstable_p_in_I_kq_in_II} and \ref{fig:app_unstable_pk_in_II_q_in_III} show the results of numerical simulations of the system~(\ref{e:triad_dynamical_system}) when it is forced at intermediate and small scales (figure~\ref{fig:app_unstable_p_in_I_kq_in_II}) and at large and intermediate scales (figure~\ref{fig:app_unstable_pk_in_II_q_in_III}). In both cases, even though the intermediate scale is forced, it is eventually suppressed and the asymptotic behaviour becomes identical to the single-mode forcing case, as described in section~\ref{sec:unstable_triads}.

\section{Fixed points of the active triadic system and their linear stability}\label{app:fixed_point}
We show that the triadic system~(\ref{e:triad_dynamical_system}) forced at the intermediate wavenumber exhibits a linearly unstable fixed point. To this end, we first look for time-independent solutions of~(\ref{e:Aeom}), satisfying
\be
\mathsfbi D\mathsfbi I\bs A&=&2\Delta (\mathsfbi I\bs A^* \times \bs A^*).
\ee
Remembering the convention $p<k<q$ and using the polar representation $A_k=|A_k| e^{i \phi_k}$ we demand that
\be
-s_k|D_{kk}|k^2|A_k| e^{i\phi}&=&2\Delta |p^2-q^2||A_p||A_q|, \\
s_p|D_{pp}|p^2|A_p| e^{i\phi}&=&2\Delta |q^2-k^2||A_q| |A_k|, \\
s_q|D_{qq}|q^2|A_q| e^{i\phi}&=&2\Delta |k^2-p^2||A_k| |A_p|,
\ee
where $s_k=1$ if $D_{kk}>0$ and $s_k=-1$ if $D_{kk}<0$ and $\phi=\phi_k+\phi_p+\phi_q$. Matching the phases, requires that
\be
\phi+\phi_{-s_k}=\phi+\phi_{s_p}=\phi+\phi_{s_q}=0,
\ee
where the equalities hold modulo $2\pi$. The only way to satisfy the above restrictions is to choose $s_k=-1$ and $s_p=s_q=1$, that is,  a fixed point can exist only when the intermediate wavenumber is forced. Of course, we must then have $\phi=0$, which leaves a two-parameter family of fixed points. Without loss of generality, we can set all phases to zero $\phi_k=\phi_p=\phi_q=0$. Matching the amplitudes gives
\be
\label{e:app_amplitudes}
|D_{kk}|k^2|A_k|&=&2\Delta |p^2-q^2||A_p||A_q|, \\
|D_{pp}|p^2|A_p|&=&2\Delta |q^2-k^2||A_q| |A_k|, \\
|D_{qq}|q^2|A_q|&=&2\Delta |k^2-p^2||A_k| |A_p|.
\ee
Furthermore, we still have the following two identities
\be
-|D_{kk}|k^2|A_k|^2+|D_{pp}|p^2|A_p|^2+|D_{qq}|q^2|A_q|^2&=&0, \\
-|D_{kk}|k^4|A_k|^2+|D_{pp}|p^4|A_p|^2+|D_{qq}|q^4|A_q|^2&=&0,
\ee
which represent energy and in-plane enstrophy  balance: energy  and enstrophy produced at the wavenumber $k$ are dissipated at wavenumbers $p$ and $q$. The two constraints leave one degree of freedom represented by the line
\be
\left[
\begin{array}{c}
     |A_k|^2        \\ 
     |A_p|^2       \\
     |A_q|^2
\end{array}     
\right]
     = \alpha
\left[
\begin{array}{ccc}     
        \f{|p^2-q^2|}{|D_{kk}|k^2}\\[0.3em]
     \f{|q^2-k^2|}{|D_{pp}|p^2} 	\\[0.3em]
     \f{|k^2-p^2|}{|D_{qq}|q^2} 
\end{array}       
\right].  
\ee
The positive constant $\alpha$ is fixed by inserting the above expression into~(\ref{e:app_amplitudes}), which then yields  for $\bs A$
the fixed point\be
\label{e:Afixed_point}
\left[
\begin{array}{c}
     A_k        \\ 
     A_p       \\
     A_q
\end{array}     
\right]
     =
     \alpha^{1/2}
\left[
\begin{array}{ccc}     
          \sqrt{\f{|p^2-q^2|}{|D_{kk}|k^2}}  \\[0.3em]
     \sqrt{\f{|q^2-k^2|}{|D_{pp}|p^2}} 	\\[0.3em]
     \sqrt{\f{|k^2-p^2|}{|D_{qq}|q^2}} 
\end{array}       
\right],
\ee
 where
\be
\alpha=-\det(\mathsfbi I\mathsfbi D)/(4\Delta^2|p^2-q^2||q^2-k^2||k^2-p^2|).
\ee
All other fixed points are obtained by the transformation
\be
\label{e:app_fixed_point_phase_shift}
\left[
\begin{array}{c}
     A_k        \\ 
     A_p       \\
     A_q
\end{array}     
\right]
\to
\left[
\begin{array}{c}
     A_k e^{i\phi_k}       \\ 
     A_p e^{i\phi_p}      \\
     A_q e^{i\phi_q}
\end{array}     
\right]
\ee
where $\phi_k+\phi_p+\phi_q=0$. 
\par
We now turn to the fixed points of the system for $\bs B(t)= \Real\bs B(t)+i \Imag \bs B(t)$, that is, we look for time-independent solutions of~(\ref{e:Beom}) with $\bs A$ given by~(\ref{e:Afixed_point}). In this case, the system decouples into two linear equations for the real and imaginary parts 
\be
\mathsfbi D\Real \bs B &=&2\Delta\Real\bs B \times\bs A, \\
\mathsfbi D\Imag \bs B &=&-2\Delta\Imag\bs B \times\bs A.
\ee 
In both cases the null-space is one dimensional, generated by the vectors
\be\label{e:fixB}
      \left[
      \begin{array}{c}
     \Real B_k           \\
     \Real B_p 		\\
     \Real B_q
     \end{array}
    \right]  
     =     
      \left[
      \begin{array}{c}
     k \sqrt{\f{|p^2-q^2|}{|D_{kk}|}}           \\
     p \sqrt{\f{|q^2-k^2|}{|D_{pp}|}} 		\\
     q \sqrt{\f{|k^2-p^2|}{|D_{qq}|}} 
     \end{array}
     \right],
     \qquad
           \left[
      \begin{array}{c}
     \Imag B_k           \\[0.3em]
     \Imag B_p 		\\[0.3em]
     \Imag B_q
     \end{array}
    \right]     
     =     
      \left[
      \begin{array}{c}
     k \sqrt{\f{|p^2-q^2|}{|D_{kk}|}}(-k^2+p^2+q^2)           \\
     p \sqrt{\f{|q^2-k^2|}{|D_{pp}|}} (k^2-p^2+q^2)		\\
     q \sqrt{\f{|k^2-p^2|}{|D_{qq}|}} (k^2+p^2-q^2)
     \end{array}
    \right]
    \qquad
\ee
The fixed point for $\bs B$ is obtained by combining the real and imaginary parts,
\be
\left[
\begin{array}{c}
     B_k           \\
     B_p 		\\
     B_q
\end{array}     
\right]   
     &=&  
\alpha^{1/2}
\left[
\begin{array}{ccc}
     k \sqrt{\f{|p^2-q^2|}{|D_{kk}|}} [c_1+ic_2(-k^2+p^2+q^2)]           \\[0.3em]
     p \sqrt{\f{|q^2-k^2|}{|D_{pp}|}} [c_1+ic_2(k^2-p^2+q^2)]		\\[0.3em]
     q \sqrt{\f{|k^2-p^2|}{|D_{qq}|}} [c_1+ic_2(k^2+p^2-q^2)]
\end{array}     
\right],
\ee
where $c_1$ and $c_2$ are some arbitrary real constants and the prefactor $\alpha^{1/2}$ has been factored out for convenience. Note that if we started with any other fixed point for $\bs A$ obtained by the transformation~(\ref{e:app_fixed_point_phase_shift}), then the above argument still applies, provided we apply the same phase transformation to the vector $\bs B$.
The real constants $c_1$ and $c_2$ set the helicity and energy of the fixed point. Indeed
\be
H=2\mathsfbi I\bs A\cdot \Real\bs B=
2\alpha c_1 \Big(k^2\f{|p^2-q^2|}{|D_{kk}|}+p^2\f{|q^2-k^2|}{|D_{pp}|}+q^2\f{|k^2-p^2|}{|D_{qq}|}\Big),
\ee
and
\be
\f{2E}{\alpha}&=&\f{|p^2-q^2|}{|D_{kk}|}+\f{|q^2-k^2|}{|D_{pp}|}+\f{|k^2-p^2|}{|D_{qq}|} +\nonumber\\ 
&&
 c_1^2 \Big(k^2\f{|p^2-q^2|}{|D_{kk}|}+p^2\f{|q^2-k^2|}{|D_{pp}|}+q^2\f{|k^2-p^2|}{|D_{qq}|}\Big) +\nonumber\\
&&
 c_2^2 \Big(k^2\f{|p^2-q^2|}{|D_{kk}|}(p^2+q^2-k^2)^2+p^2\f{|q^2-k^2|}{|D_{pp}|}(q^2+k^2-p^2)^2 +\nonumber\\
 &&\,\,\,\,\,\,\,q^2\f{|k^2-p^2|}{|D_{qq}|}(k^2+p^2-q^2)^2\Big).
\ee
We now show that the fixed point for the triadic system (\ref{e:triad_dynamical_system}) is linearly unstable by studying the perturbation $\bs A=\bar{\bs A}+\delta \bs A$ around the fixed point $\bar{\bs A}$ given by~(\ref{e:Afixed_point}). The real and imaginary parts of the linearized dynamical equation~(\ref{e:Aeom}) for $\delta\bs A=\Real\delta\bs A+i\Imag \delta\bs A$ read
\be
\Real\delta\dot{\bs A}+\mathsfbi D\Real\delta\bs A&=&
2\Delta \mathsfbi I^{-1}(\mathsfbi I\Real\delta\bs A \times \bar{\bs A})+
2\Delta \mathsfbi I^{-1}(\mathsfbi I\bar{\bs A} \times \Real\delta\bs A), \\
\Imag\delta\dot{\bs A} +\mathsfbi D\Imag\delta\bs A&=&
-2\Delta \mathsfbi I^{-1}(\mathsfbi I\Imag\delta\bs A \times \bar{\bs A})-
2\Delta \mathsfbi I^{-1}(\mathsfbi I\bar{\bs A} \times \Imag\delta\bs A).
\ee
Since these two equations are decoupled, it suffices to show linear instability of the first equation. The corresponding  Jacobian $\mathsfbi J$ reads 
\be
\mathsfbi J=-\mathsfbi D-2\Delta \mathsfbi I^{-1}\mathsfbi M_{\bar{\bs A}}\mathsfbi I+2\Delta I^{-1}\mathsfbi M_{\mathsfbi I\bar{\bs A}},
\ee
where $\mathsfbi M_{\bs w}$ denotes the antisymmetric matrix with components $M_{ab}=\eps_{acb} w_c$, corresponding to the cross product with $\bs w$. Direct computation reveals that the Jacobian has the following properties
\be
\label{e:AR_Jacobian_properties}
\tn{tr}{(\mathsfbi J)}=-\tn{tr}{(\mathsfbi D)}, \quad \tn{tr}(\mathsfbi J^2)-\tn{tr}^2(\mathsfbi J)=0,\quad \det{(\mathsfbi J)}=4\det{(\mathsfbi D)}.
\ee
We recall the Routh-Hurwitz stability criteria for the eigenvalues of a $3\times 3$ matrix $\mathsfbi M$ to have negative real parts~\citep{gantmakher1998theory}
\be
\tn{tr}{(\mathsfbi M)}<0, \qquad 
\det{(\mathsfbi M)}<0, \qquad 
\tn{tr}{(\mathsfbi M)}[\tn{tr}(\mathsfbi M^2)-\tn{tr}^2(\mathsfbi M)] >-2\det{(\mathsfbi M)}.
\ee
The Jacobian $\mathsfbi J$ satisfies the first condition because of our restriction~(\ref{e:pos_trace}), it also satisfies the second condition because the fixed point only exists for $\det \mathsfbi D<0$. But it violates the last one, since for the fixed point one always has $\det \mathsfbi D<0$. Thus  $\mathsfbi J$ has an eigenvalue with positive or vanishing real part. We now show that the real part is always positive, implying that the fixed point is linearly unstable. To this end, note that the properties~(\ref{e:AR_Jacobian_properties}) imply that the characteristic equation of $\mathsfbi J$ has the form
\be
\label{e:Jacobian_char_eq}
\det(\lambda\mathsfbi I-\mathsfbi J)=\lambda^3+\lambda^2\tn{tr}(\mathsfbi D)+4|\det{(\mathsfbi D)}|=0.
\ee
Since we assume that $\tn{tr}(\mathsfbi D)>0$, this cubic equation has negative discriminant
\be
-16\tn{tr}^3(\mathsfbi D) |\det{(\mathsfbi D)}|-432|\det{(\mathsfbi D)}|^2<0,
\ee
implying that~(\ref{e:Jacobian_char_eq}) has one real root and two non-real complex conjugate roots. Equivalently, (\ref{e:Jacobian_char_eq}) must have the form
\be
\label{e:char_eq_neg_discr}
\det(\lambda\mathsfbi I-\mathsfbi J)=(\lambda-r_1)(\lambda-r_2)(\lambda-r_2^*),
\ee
where $r_1$ is real and $r_2$ is complex. Thus, we want to eliminate the possibility that $\Real(r_1)=r_1=0$ or $\Real(r_2)=0$. If $r_1=0$, then~(\ref{e:char_eq_neg_discr}) reduces to
\be
\det(\lambda\mathsfbi I-\mathsfbi J)=\lambda(\lambda-r_2)(\lambda-r_2^*)=\lambda^3-\lambda^2(r_2+r_2^*)+\lambda|r_2|^2,
\ee
which is incompatible with (\ref{e:Jacobian_char_eq}), since $|\det{(\mathsfbi D)}|\neq 0$ for the active triads considered here.  If $\Real(r_2)=0$, then~(\ref{e:char_eq_neg_discr}) reduces to, for some real $r$,
\be
\det(\lambda\mathsfbi I-\mathsfbi J)=(\lambda-r_1)(\lambda-ir)(\lambda+ir)=\lambda^3-\lambda^2 r_1+\lambda r^2-r_1r^2,
\ee
which is also incompatible with (\ref{e:Jacobian_char_eq}), since imposing that $r=0$ to eliminate the term proportional to $\lambda$, also eliminates the constant term. Thus, $\mathsfbi J$ has at least one eigenvalue with positive real part, implying that the fixed point~(\ref{e:triad_dynamical_system}) is linearly unstable.

\section{The phase space of the system for $\bs A(t)$ when $\mathsfbi D= \mathsfbi 0$}\label{app:torus}

\subsection{Geometry of the solutions}

Consider the system~(\ref{e:Aeom}) when $\mathsfbi D= \mathsfbi 0$
\begin{eqnarray}
    \label{e:classical_A}
    \left. \begin{array}{ll}  
        k^2\dot{A}_k=2\Delta(p^2-q^2)A_p^*A_q^*\\[8pt]
        p^2\dot{A}_p=2\Delta(q^2-k^2)A_q^*A_k^* \\[8pt]
        q^2\dot{A}_q=2\Delta(k^2-p^2)A_k^*A_p^*
    \end{array}\right\},
\end{eqnarray}
which has the three constants of motion
\begin{eqnarray}
    \label{e:three_constants}
    \left. \begin{array}{r}  
        k^2 |A_k|^2+p^2 |A_p|^2+q^2 |A_q|^2=E \\
        k^4 |A_k|^2+p^4 |A_p|^2+q^4 |A_q|^2=\Omega \\
        A_kA_pA_q-A_k^*A_p^*A_q^*=C
    \end{array}\right\}.
\end{eqnarray}
The quadratic invariants $E$ and $\Omega$ were found by~\citet{moffatt2014note}, and 
the cubic invariant~$C$ was derived in section~\ref{sec:cubic_inv}.

Equations~(\ref{e:three_constants}) provide three constraints 
for $(A_k,A_p,A_q) \in \mathbb{C}^3 \simeq \mathbb{R}^6$
depending on
$(E,\Omega,C) \in \mathbb{R}_{\geq 0} \times \mathbb{R}_{\geq 0} \times i \mathbb{R}$.
Denote by $M_{(E,\Omega,C)}$ the set defined by these equations.
We will show that for generic values of $(E,\Omega,C)$
the set $M_{(E,\Omega,C)}$ is a~compact three-dimensional manifold
(possibly empty) and that each of its connected components is a~three-torus.

To show that the system~(\ref{e:three_constants}) defines a manifold 
in an appropriate subset of $\mathbb R^6$, 
it is enough to show that its differential $\mathsfbi J$ has full rank on that subset. 
Differentiating~(\ref{e:three_constants}) with respect 
to $\partial_{A_i}$ and $\partial_{A_i^*}$ yields
\begin{eqnarray}
    \mathsfbi J=
    \left[
        \begin{array}{llllll}
            k^2A_k^* & k^2A_k      & p^2A_p^* & p^2A_p      & q^2A_q^* & q^2A_q       \\ 
            k^4A_k^* & k^4A_k      & p^4A_p^* & p^4A_p      & q^4A_q^* & q^4A_q       \\
            A_pA_q   & -A_p^*A_q^* & A_kA_q   & -A_k^*A_q^* & A_kA_p   & -A_k^*A_p^*
        \end{array}     
    \right].
\end{eqnarray}
Note that the matrix above is in fact the complexification of $\mathsfbi J$,
which has the same rank.
First, consider the minor $J_{123}$:
\begin{eqnarray}
    J_{123}=\det
    \left[
        \begin{array}{lll}
            k^2A_k^* & k^2A_k      & p^2A_p^*     \\ 
            k^4A_k^* & k^4A_k      & p^4A_p^*     \\
            A_pA_q   & -A_p^*A_q^* & A_kA_q 
        \end{array}    
    \right]=
    2p^2k^2(p^2-k^2)A_p^*\Real(A_kA_pA_q).
\end{eqnarray}
We see that $\Real(A_kA_pA_q)\neq 0$ implies that $\mathsfbi J$ has full rank. 
We now consider the various cases when $\Real(A_kA_pA_q)=0$.

\textbf{Case 1}. 
Two (or more) modes vanish, say $A_p=A_q=0$. 
Then the last row of $\mathsfbi J$ is zero 
and thus $\mathsfbi J$ can have rank at most 2. 
Therefore, we will consider the subset
\begin{eqnarray}
    \label{e:M1subset}
    Z_1=\{\, A_p=0, A_q=0\}\cup \{\, A_q=0,  A_k=0\} \cup \{\, A_k=0,  A_p=0\}
\end{eqnarray}
of $\mathbb{R}^6$ separately.

\textbf{Case 2}.
One mode vanishes, say $A_k=0$ but $A_p\neq 0$ and $A_q \neq 0$. 
The differential $\mathsfbi J$ takes the form
\begin{eqnarray}
    \mathsfbi J|_{A_k=0}=
    \left[
        \begin{array}{cccccc}
            0      & 0           & p^2A_p^* & p^2A_p & q^2A_q^* & q^2A_q    \\ 
            0      & 0           & p^4A_p^* & p^4A_p & q^4A_q^* & q^4A_q    \\
            A_pA_q & -A_p^*A_q^* & 0 & 0    & 0 & 0
        \end{array}     
    \right].
\end{eqnarray}
Taking linear combination of the first two rows gives
\begin{eqnarray}
    \mathsfbi{\tilde J}|_{A_k=0}=
    \left[
        \begin{array}{cccccc}
            0      & 0           & 0        & 0      & q^2A_q^*(p^2-q^2) & q^2A_q (p^2-q^2) \\ 
            0      & 0           & p^4A_p^* & p^4A_p & q^4A_q^*          & q^4A_q           \\
            A_pA_q & -A_p^*A_q^* & 0        & 0      & 0                 & 0
        \end{array}     
    \right],
\end{eqnarray}
which has full rank, since $A_p\neq 0$ and $A_q \neq 0$.

\textbf{Case 3}.
None of the modes vanish, i.e. $|A_k||A_p||A_q|\neq 0$, but 
$\mathrm{Re}(A_kA_pA_q) = 0$. To simplify the analysis, note that the system~(\ref{e:three_constants}) has the property (ii) of section~\ref{sec:cubic_inv}, that is, it is invariant under the change of variables
\begin{eqnarray}
    (A'_k, A'_p, A'_q) = (A_k e^{i\psi_k}, A_p e^{i\psi_p}, A_q e^{i\psi_q}) 
    \quad \tn{where } \psi_k + \psi_p + \psi_q = 0.
\end{eqnarray}
 Therefore, without loss of generality we can assume
$A_p \in \mathbb{R}, A_q \in \mathbb{R}$,
and then $\mathrm{Re}(A_kA_pA_q) = 0$ together with $|A_kA_pA_q| \neq 0$
implies $A_k \in i\mathbb{R}$.
The differential becomes
\begin{eqnarray}
    \mathsfbi J=
    \left[
        \begin{array}{llllll}
            -k^2A_k & k^2A_k   & p^2A_p & p^2A_p & q^2A_q & q^2A_q    \\ 
            -k^4A_k & k^4A_k   & p^4A_p & p^4A_p & q^4A_q & q^4A_q    \\
            A_p A_q & -A_p A_q & A_kA_q & A_kA_q & A_kA_p & A_kA_p
        \end{array}     
    \right].
\end{eqnarray}
The second, fourth and sixth columns are, up to a sign, 
the same as the first, third and fifth columns, respectively. 
Thus $\mathsfbi J$ has full rank if and only if the minor $J_{135}$ is nonzero.
We have:
\begin{eqnarray}
    J_{135}&=&
    \det
    \left[
        \begin{array}{lll}
            -k^2A_k & p^2A_p & q^2A_q       \\ 
            -k^4A_k & p^4A_p & q^4A_q    \\
            A_pA_q  & A_kA_q & A_kA_p 
        \end{array}    
    \right] \nonumber\\
    & = &
    - A_q^2 A_k^2 k^2q^2 (k^2-q^2) 
    - A_p^2 A_k^2 p^2k^2 (p^2-k^2) 
    + A_p^2 A_q^2 q^2p^2 (q^2-p^2) \nonumber\\
    & = &
      |A_q|^2 |A_k|^2 k^2 q^2 (k^2 - q^2) 
    + |A_p|^2 |A_k|^2 p^2 k^2 (p^2 - k^2)+ \nonumber\\
    &&|A_p|^2 |A_q|^2 q^2 p^2 (q^2 - p^2).
\end{eqnarray}
Therefore, we must treat the following subset separately:
\begin{eqnarray}
    \label{e:M3subset}
    Z_3 = && \{\,|A_k||A_p||A_q|\neq 0, 
        \mathrm{Re}(A_kA_pA_q) = 0\}\cap \nonumber\\
    &&  \{\, 
      |A_q|^2 |A_k|^2 k^2 q^2 (k^2 - q^2) 
    + |A_p|^2 |A_k|^2 p^2 k^2 (p^2 - k^2)+\nonumber\\
    && \;\; |A_p|^2 |A_q|^2 q^2 p^2 (q^2 - p^2)
    = 0 \},
\end{eqnarray}
which will be analysed in section~\ref{sec:exac_sol_z1z3}.

We conclude that the system~(\ref{e:three_constants}) defines 
a~foliation of $N = \mathbb{R}^6 \setminus (Z_1 \cup Z_3)$ by three-dimensional manifolds
since the differential $\mathsfbi J$ has full rank on $N$.
Precisely, $N$ is foliated by the manifolds $\tilde M_{(E,\Omega,C)} = M_{(E,\Omega,C)} \cap N$.
We call the closed set
\begin{eqnarray}
    Z & = & \{ \mathrm{rk} \mathsfbi J < 3 \} =
    Z_1\cup Z_3 = \{\Real(A_kA_pA_q)=0\}  \cap\\
    &&  \{
      |A_q|^2 |A_k|^2 k^2 q^2 (k^2 - q^2) 
    + |A_p|^2 |A_k|^2 p^2 k^2 (p^2 - k^2) 
    + |A_p|^2 |A_q|^2 q^2 p^2 (q^2 - p^2)
    =0\},\nonumber
\end{eqnarray}
the \emph{singular locus} (of $\mathsfbi J$).
Its complement, $N = \mathbb{R}^6 \setminus Z$, is called the 
\emph{regular locus} (of $\mathsfbi J$).

The considerations above prove that $\tilde M_{(E,\Omega,C)}$ is a~three-dimensional 
smooth submanifold of $\mathbb{R}^6$.
We now prove that for generic values of $(E,\Omega,C)$ the set $M_{(E,\Omega,C)}$
does not intersect $Z$ and thus is equal to $\tilde M_{(E,\Omega,C)}$,
and therefore is a~compact three-dimensional submanifold.
Moreover, we prove that it is in fact a~sum of disjoint copies of
the three-torus $\mathbb{T}^3$.

First, note that $M_{(E,\Omega,C)}$, 
as well as the sets $Z_1$ and $Z_3$ are invariant under the change of variables
\begin{eqnarray}
    (A_k,A_p,A_q)
    \mapsto (A_k e^{i\psi_k},A_p e^{i\psi_p},A_q e^{i\psi_q}) 
    \quad \tn{where } \psi_k+\psi_p+\psi_q=0
\end{eqnarray}
which defines a group action of the two-torus 
$\mathbb{T}^2 = S^1 \times S^1$ on $M_{(E,\Omega,C)}$.
For $g \in \mathbb{T}^2$,
denote by $g \cdot x$ the action of the group element $g$ on $x$.
Moreover, this action is free on $\mathbb{R}^6 \setminus Z_1$,
and in particular on every $\tilde M_{(E,\Omega,C)}$.
By Corollary 21.6 and Theorem 21.10 in~\citep{lee2013manifolds}
the orbit space $\tilde O_{(E,\Omega,C)} = \tilde M_{(E,\Omega,C)}/\mathbb{T}^2$ 
is a smooth manifold of dimension~$1$.
Thus, the manifold $\tilde M_{(E,\Omega,C)}$
is a~fiber bundle over $\tilde O_{(E,\Omega,C)}$
with fiber $\mathbb{T}^2$
(in fact, it is a~principal $\mathbb{T}^2$-bundle).
We denote the quotient map by 
$\Pi: \tilde M_{(E,\Omega,C)} \to \tilde O_{(E,\Omega,C)}$.

Since $\tilde O_{(E,\Omega,C)}$ is $1$-dimensional,
it is a~union of circles $S^1$ and lines $\mathbb{R}$.
Consider any component $\tilde O$ of $\tilde O_{(E,\Omega,C)}$
and the component $\tilde M$ of $\tilde M_{(E,\Omega,C)}$
projecting to $\tilde O$, i.e. $\tilde M = \Pi^{-1}(\tilde O)$.
Suppose $\tilde O$ is diffeomorphic to $\mathbb{R}$.
Since $\mathbb{R}$ is contractible,
every fiber bundle over it is trivial,
so $\tilde M$ is diffeomorphic to $\mathbb{R} \times \mathbb{T}^2$.

Suppose now $\tilde O$ is diffeomorphic to $S^1$.
Consider the map $\gamma:[0,1] \to \tilde O \simeq S^1$ 
given by $\gamma(t) = e^{2 \pi i t}$.
Lift this map to a~map $\bar \gamma:[0,1] \to \tilde M$,
that is, take any map such that $\Pi(\bar \gamma(t)) = \gamma(t)$.
Note that $\gamma(0) = \gamma(1) = 1$,
thus $\bar \gamma(0),\bar \gamma(1)$ belong to the same fiber of $\Pi$,
$\Pi^{-1}(1)$.
But $\mathbb{T}^2$ acts transitively on the fibers of $\Pi$,
thus there is an element $g \in \mathbb{T}^2$ such that
$\bar \gamma(0) = g \cdot \bar \gamma(1)$.
Now take a~path $s:[0,1] \to \mathbb{T}^2$
such that $s(1)=g$ is as above and $s(0)$
is the identity element.
Then the map $\eta(t) = s(t) \bar \gamma(t)$ has the property that
$\eta(0) = \eta(1)$, and thus it descends to a~map $\eta:S^1 \to \mathbb{T}^2$
such which lifts $\gamma$, i.e. $\Pi \circ \eta = \gamma$.
Finally, after smoothing $\eta$, the map
$F:\mathbb{T}^3 = S^1 \times \mathbb{T}^2 \to \tilde M$
given by $F(t,g) = g \cdot \eta(t)$
gives the desired diffeomorphism of $\mathbb{T}^3$ and $\tilde M$.

In particular, what follows is that whenever $M_{(E,\Omega,C)}$ does not intersect $Z$,
it is a~disjoint union of a~finite number of three-tori.
This may be empty when $M_{(E,\Omega,C)}$ is empty, for instance if 
$\Omega/E > \max(k^2,p^2,q^2)$, or $\Omega/E < \min(k^2,p^2,q^2)$ etc.
Now we determine a~residual subset of triples 
$(E,\Omega,C) \in \mathbb{R}_{\geq 0} \times \mathbb{R}_{\geq 0} \times i\mathbb{R}$
for which $M_{(E,\Omega,C)} \cap Z = \varnothing$.

Consider $(A_k,A_p,A_q) \in M_{(E,\Omega,C)} \cap Z$.
Since $\mathrm{Re}(A_kA_pA_q) = 0$,
we have $C = A_kA_pA_q - A_k^*A_p^*A_q^*
= 2 i \mathrm{Im}(A_kA_pA_q) = 2 A_k A_p A_q$,
thus $|C|^2 = 4 |A_k|^2 |A_p|^2 |A_q|^2$.
Denote $x = |A_k|^2, y = |A_p|^2, z = |A_q|^2$.
The system (\ref{e:three_constants}) together with the equations defining $Z$ 
thus implies
\begin{eqnarray}
    \left. \begin{array}{r}
        k^2 x + p^2 y + q^2 z = E \\
        k^4 x + p^4 y + q^4 z = \Omega \\
          k^2 q^2 (k^2 - q^2) x z 
        + p^2 k^2 (p^2 - k^2) x y
        + q^2 p^2 (q^2 - p^2) y z
        = 0 \\
        4xyz = |C|^2
    \end{array}
    \label{e:modulus-equations-Z} \right\}.
\end{eqnarray}
The first two equations express $y, z$ as linear functions of $x$.
Inserting these into the third equation one obtains a~quadratic equation for $x$ with a non-zero leading term,
which has at most $2$ solutions.
These solutions give at most $2$ possible values of $|C|$ using the last equation.
Denote the set of triples $(E,\Omega,C)$ obtained this way by $S$.
This is a~codimension $1$ subset, 
thus a~generic $(E,\Omega,C)$ does not belong to $S$,
and for such a~triple $(E,\Omega,C)$ outside of $S$
the set $M_{(E,\Omega,C)}$ is deemed to be a~sum of three-tori as explained earlier.

Since the differential $\mathsfbi J$ is of full rank on $N = \mathbb{R}^6 \setminus Z$,
the set $F^{-1}(S) \cap N$ is of codimension $1$, where $F:  \mathbb{R}^6\to  \mathbb{R}^3$ is the map determined by~(\ref{e:three_constants}).
However, the set $Z$ is of codimension $1$, too,
and since $F^{-1}(S) = (F^{-1}(S) \cap N) \cup (F^{-1}(S) \cap Z)$,
we conclude that $F^{-1}(S)$ is of codimension $1$.
The complement of this set is foliated by three-tori,
so taking all things together it follows that a~generic point in
$\mathbb{R}^6$ lies on one of these smooth three-tori.

\subsection{Exact solutions for initial conditions on $Z_1$ and $Z_3$}\label{sec:exac_sol_z1z3}

To finish this section, we comment on the nature of the dynamics~(\ref{e:classical_A}) 
when the initial conditions are taken from the subsets $Z_1$ and $Z_3$. 
It is easy to see that points on $Z_1$ are simply fixed points. 
Taking initial conditions on $Z_3$ results in evolution with constant amplitudes 
$|A_k|$, $|A_p|$ and $|A_q|$ and phases exhibiting periodic motion on two-torus. 
Indeed, consider the following ansatz 
$(A_k,A_p,A_q)=(|A_k|e^{i\phi_k(t)},|A_p|e^{i\phi_p(t)},|A_q|e^{i\phi_q(t)})$, 
where the amplitudes are independent of time. 
The system~(\ref{e:classical_A}) gives
\begin{eqnarray}\label{e:ansatz-case3}
    \left. \begin{array}{ll}  
        k^2|A_k|\dot{\phi}_k = (p^2-q^2)|A_p||A_q|e^{-i(\phi_k+\phi_p+\phi_q+\pi/2)}\\[8pt]
        p^2|A_p|\dot{\phi}_p = (q^2-k^2)|A_q||A_k| e^{-i(\phi_k+\phi_p+\phi_q+\pi/2)}\\[8pt]
        q^2|A_q|\dot{\phi}_q = (k^2-p^2)|A_k||A_p|e^{-i(\phi_k+\phi_p+\phi_q+\pi/2)}
    \end{array}\right\}.
\end{eqnarray}
On $Z_3$, $e^{-i(\phi_k+\phi_p+\phi_q+\pi/2)}=\pm 1$.
If we assume that this holds for any time $t$,
we easily find solutions to these equations:
\begin{eqnarray}\label{e:solution-case3}
    \left. \begin{array}{ll}  
        \phi_k = \pm(p^2-q^2) |A_p||A_q| / (k^2|A_k|) t + c_k\\[8pt]
        \phi_p = \pm(q^2-k^2) |A_q||A_k| / (p^2|A_p|) t + c_p\\[8pt]
        \phi_q = \pm(k^2-p^2) |A_k||A_p| / (q^2|A_q|) t + c_q
    \end{array} \right\}.
\end{eqnarray}
Moreover, multiplying the first equation of (\ref{e:ansatz-case3})
by $p^2q^2|A_p||A_q|$, 
the second by $k^2q^2|A_k||A_q|$, 
the third by $k^2p^2|A_k||A_p|$ and adding them together gives
\begin{eqnarray}
    \pm k^2p^2q^2 |A_k||A_p||A_q| (\dot{\phi}_k+\dot{\phi}_p+\dot{\phi}_q)
    & = & (p^2-q^2)p^2q^2 |A_p|^2|A_q|^2+ \nonumber \\
    &  & (q^2-k^2)q^2k^2 |A_q|^2|A_k|^2+ \nonumber \\
    &  & (k^2-p^2)k^2p^2 |A_k|^2|A_p|^2\nonumber \\
    & = & 0.
\end{eqnarray}
The right-hand side is zero by the definition of $Z_3$, 
implying that the sum of phases $\phi_k+\phi_p+\phi_q$ is indeed constant 
and equal to $\pi/2$ or $3\pi/2$ also by the definition of $Z_3$.
Therefore (\ref{e:solution-case3}) gives the solutions to the system (\ref{e:classical_A})
on $Z_3$ and these exhibit quasi-periodic motion.


\end{document}